\documentclass[12pt]{article}

\usepackage{amsmath}
\usepackage{amssymb}
\usepackage{amsfonts}
\usepackage{graphicx}  
\usepackage[utf8]{inputenc}		
\usepackage{aas_macros}

\usepackage[dvipsnames]{xcolor}
\usepackage{fancyhdr}		
\usepackage{lipsum}			
\usepackage{framed}			
\usepackage{cite}			

\usepackage{listings}      

\DeclareFixedFont{\ttb}{T1}{txtt}{bx}{n}{12} 
\DeclareFixedFont{\ttm}{T1}{txtt}{m}{n}{12}  

\usepackage{color}
\definecolor{deepblue}{rgb}{0,0,0.5}
\definecolor{deepred}{rgb}{0.6,0,0}
\definecolor{deepgreen}{rgb}{0,0.5,0}

\newcommand\pythonstyle{\lstset{
language=Python,
basicstyle=\ttm,
otherkeywords={self},             
keywordstyle=\ttb\color{deepblue},
emph={MyClass,__init__},          
emphstyle=\ttb\color{deepred},    
stringstyle=\color{deepgreen},
frame=tb,                         
showstringspaces=false            %
}}

\lstnewenvironment{python}[1][]
{
\pythonstyle
\lstset{#1}
}
{}

\newcommand\pythoninline[1]{{\pythonstyle\lstinline!#1!}}

\usepackage{booktabs}		

\usepackage[font=small]{caption} 
\usepackage{float}         
 
\usepackage[margin=2cm]{geometry}   
\graphicspath{{figures/}}			
\numberwithin{equation}{section}    

\usepackage{multicol}
\usepackage{etoolbox}
\usepackage{relsize}
\patchcmd{\thebibliography}
  {\list}
  {\begin{multicols}{2}\smaller\list}
  {}
  {}
\appto{\endthebibliography}{\end{multicols}}

\let\oldenumerate\enumerate
\renewcommand{\enumerate}{
  \oldenumerate
  \setlength{\itemsep}{1pt}
  \setlength{\parskip}{0pt}
  \setlength{\parsep}{0pt}
}

\let\olditemize\itemize
\renewcommand{\itemize}{
  \olditemize
  \setlength{\itemsep}{4pt}
  \setlength{\parskip}{0pt}
  \setlength{\parsep}{0pt}
}

\newcommand{\email}[1]{\href{mailto:#1}{#1}}
\newenvironment{institutions}[1][2em]{\begin{list}{}{\setlength\leftmargin{#1}\setlength\rightmargin{#1}}\item[]}{\end{list}}

\definecolor{myblue1}{RGB}{76, 142, 185}
\definecolor{myblue2}{RGB}{25, 100, 126}
\definecolor{myblue3}{RGB}{41, 110, 180}

\definecolor{mygreen1}{RGB}{88,165,87}
\definecolor{mygreen2}{RGB}{91,165,98}

\definecolor{myred1}{RGB}{221, 28, 26}
\definecolor{mypurple}{RGB}{122,48,108}

\usepackage{xhfill}

\usepackage{dirtree}

\usepackage{subfig}

\usepackage{wrapfig}
\usepackage{siunitx}
\usepackage{tikz}
\usetikzlibrary{shapes, arrows}

\usepackage[
	colorlinks=true,
	citecolor=green!50!black,
	linkcolor=NavyBlue!75!black,
	urlcolor=green!50!black,
	hypertexnames=false]{hyperref}

\usepackage{cleveref}
\crefname{equation}{Eq.}{Eqs.}
\crefname{figure}{Fig.}{Figs.}
\crefname{table}{Tab.}{Tabs.}

\fancypagestyle{firststyle}{
	\rhead{\footnotesize%
	\texttt{UCR-TR-2018-FLIP-L3-37}%
	}}

\usepackage{etoc}

\begin{document}

\thispagestyle{firststyle} 	

\begin{center}

    {\Large \bf
    	\texttt{DarkCapPy:} Dark Matter Capture and Annihilation
    }

    \vskip .7cm

    { \bf 
    	Adam Green
    	and 
    	Philip Tanedo
    	} 
    \\ 
    \vspace{-.2em}
    { \tt \footnotesize
	    \email{agree019@ucr.edu},
	    \email{flip.tanedo@ucr.edu}
    }
	
    \vspace{-.2cm}

    \begin{institutions}[2.25cm]
    \footnotesize
    {\it 
	    Department of Physics \& Astronomy, 
	    University of  California, Riverside, 
	    {CA} 92521	    
	    }    
    \end{institutions}

\end{center}


\begin{abstract}
\noindent 
\texttt{DarkCapPy} is a Python 3/Jupyter package for calculating rates associated with dark matter capture in the Earth, annihilation into light mediators, and the subsequent observable decay of the light mediators near the surface of the Earth. The package includes a calculation of the Sommerfeld enhancement at the center of the Earth and the timescale for capture--annihilation equilibrium. The code is open source and can be modified for other compact astronomical objects and mediator spins.  
\end{abstract}

\small
\setcounter{tocdepth}{2}
\tableofcontents
\normalsize



\newpage

\section{Program Summary}

\begin{small}
\noindent
{\em Program Title: DarkCapPy}\\
{\em Licensing provisions: Creative Commons by 4.0 (CC by 4.0)}\\
{\em Programming language: Python 3}
\\
{\em Nature of problem: }\\
  Calculate the rate at which dark matter accumulates and annihilates into light mediators in the Earth. Calculate the resulting number of signal events from light mediator decays at the IceCube Neutrino Observatory.\\
{\em Solution method:  }\\
	Python 3/Jupyter code that automates the calculation of intermediate steps and takes appropriate limits to do parameter scans in a computationally efficient manner.
\end{small}

\section{Introduction}


Dark matter that is captured in the the Earth may produce a remarkable signature of new physics. If the dark matter annihilates into light mediators, these particles may propagate to the surface of the Earth and decay into collimated, upward-going lepton pairs. This signal has very low background and can be observed at facilities such as the IceCube Neutrino Observatory \cite{Ahrens:2002dv}. 
This signature was recently explored in the context of a spin-1 dark photon mediator~\cite{Feng:2015hja}. Remarkably, the presence of the light mediator enhances the dark matter annihilation rate in the center of the Earth to produce an appreciable signal. This is in contrast to WIMP scenarios without mediators where the annihilation rate is too small for such signals to be viable. Signals from light mediators are sensitive to regions of dark matter--dark photon parameter space that are otherwise inaccessible to present experiments.

The formalism for dark matter capture in solar system objets was developed in the study of neutralinos that capture in the sun and subsequently annihilate into an observable flux of neutrinos~\cite{Freese:1985qw,Press:1985ug,Silk:1985ax, Krauss:1985aaa,Griest:1986yu,Gaisser:1986ha,Gould:1987ju, Gould:1987ir,Gould:1987ww,Gould:1991hx}.  
Subsequent studies also explored the case of Earth capture~\cite{Damour:1998rh,Damour:1998vg,Gould:1999je,Lundberg:2004dn, Peter:2009mi,Peter:2009mk,Peter:2009mm,Bruch:2009rp, Koushiappas:2009ee}.  
Augmenting the dark sector with low-mass mediator particles can change the phenomenology of this scenario to include charged leptons coming from the decay the long-lived mediators~\cite{Batell:2009zp, Schuster:2009au,Schuster:2009fc,Meade:2009mu}.

Feng, Smolinsky, and Tanedo showed that these light mediators play an additional, critical role for the captured dark matter population in the Earth~\cite{Feng:2015hja, Feng:2016ijc}. 
The mediators generate a long-range potential that causes a Sommerfeld enhancement of the non-relativistic dark matter annihilation rate. This, in turn, affects the balance in the rate equation between dark matter capture and annihilation in such a way that it is more likely for the Earth to have saturated its equilibrium captured dark matter capacity. At this equilibrium, the rate at which dark matter is captured is equal to the rate at which it annihilates into mediators. Previous examinations of this process for models without mediators found that the typical time it takes for the Earth to reach equilibrium are much larger than the age of the Earth. The equilibrium condition corresponds to the maximum flux of upward-going, long-lived mediators from the center of the Earth. 
In the case where the light mediator is a dark photon---an Abelian gauge boson that interacts with the Standard Model through kinetic mixing with hypercharge ~\cite{Kobzarev:1966qya, Okun:1982xi, Holdom:1985ag, Holdom:1986eq}---this `dark Earthshine' process results in signals from light mediator decays that probe regions of parameter space that are not accessible to traditional dark photon searches~\cite{Alexander:2016aln}.  

In this manuscript we present the the open-source Python 3/Jupyter package \texttt{DarkCapPy}. It provides functions that efficiently calculate the physical quantities necessary to determine the rate of visible upward-going Standard Model fermions coming from this `dark Earthshine' scenario.  
The inputs to this package are:
\begin{itemize}
	\item $m_X$, dark matter mass
	\item $m_{A'}$, dark mediator mass
	\item $\varepsilon$, kinetic mixing parameter between standard model photon and dark photon 
	\item $\alpha$, fine structure constant
	\item $\alpha_X$, dark fine structure constant
	\item Radius and density data for Earth
	\item Branching ratios for $A' \rightarrow e^+e^-$ as a function of $m_{A'}$.
\end{itemize}
This package outputs:
\begin{itemize}
	\item Capture rate
	\item Annihilation rate
	\item Sommerfeld enhancements
	\item Equilibrium time
	\item Signal rate at IceCube
	\item Plots of equilibrium time
	\item Plots of signal rate.
\end{itemize}

\section{Physics Overview}

The flux of upward-going mediators is determined by the rate of dark matter annihilation in the center of the Earth. 
These mediators are boosted with a decay length determined by their interactions with ordinary matter. With this information, one can subsequently determine the flux of observable charged leptons produced near the surface of the Earth.
Dark matter is captured when it loses sufficient kinetic energy in an elastic scattering off a nucleus in the Earth. We follow the conventions of Ref.~\cite{Feng:2015hja}.

\subsection{Relating Dark Matter Annihilation to Capture}

The amount of dark matter at the center of the Earth, $N_X$, depends on the relative rates of two competing processes, dark matter capture and annihilation, via the differential equation~\cite{Silk:1985ax} 
\begin{equation}
\frac{dN_X}{dt} = C_\text{cap} - N_X^2 C_\text{ann} \ .
\label{Eqn:populationEquation}
\end{equation}
$C_\text{cap}$ is the rate at which dark matter is captured in the gravitational potential of the Earth and $N_X^2 C_\text{ann}$ is the rate at which two captured dark matter particles annihilate. The $N_X^2$ prefactor accounts for the scaling of the annihilation rate with the number of available dark matter particles.
The rate at which captured dark matter population decreases is
\begin{equation}
\Gamma_\text{ann} = \frac{1}{2} N_X^2 C_\text{ann}	\ ,
\end{equation}
where the factor of $\frac{1}{2}$ accounts for the two dark matter particles that are removed in each annihilation.
We assume that dark matter is heavy enough, $m_X > 4 \ \text{GeV}$, that evaporation is negligible~\cite{Busoni:2013kaa}. We also ignore self interactions because they are a  negligible contribution to the capture rate in the Earth's gravitational potential~\cite{Zentner:2009is}.  

Inserting the solution to \cref{Eqn:populationEquation} relates the present dark matter annihilation rate to the dark matter capture rate,
\begin{equation}
\Gamma_\text{ann} = \frac{1}{2} C_\text{cap}\tanh^2\left( \frac{\tau_{\oplus}}{\tau} \right)
\label{Eqn:GammaAnnSoln}
\end{equation}
where $\tau = \sqrt{C_\text{cap} C_{\text{ann}}}$ is the characteristic time scale for the Earth to become saturated with dark matter. This is the point at which $\dot N_X \approx 0$. At this equilibrium time, every time two dark matter particles are captured, two also annihilate. This is the maximum rate at which mediators are produced at the center of the Earth and hence maximizes the flux of upward-going mediators. $\tau_{\oplus} \approx 4.5$~Gyr is the age of Earth, the relevant timescale over which dark matter may accumulate. Observe that $\tanh^2(x)$ is a steeply increasing function that flattens to unity at $x\approx 1$, thus the production of mediators is only efficient upon saturation, $\tau \gtrsim \tau_\oplus$.

\subsection{Benchmark Model: Dark Photon}

The capture and annihilation parameters $C_\text{cap}$ and $C_\text{ann}$ depend on the particle physics model of the mediator. We assume a dark photon that interacts through kinetic mixing with hypercharge, for which the effective Lagrangian is
\begin{align}
	\mathcal L &=
	-\frac 14 F_{\mu\nu}F^{\mu\nu}
	-\frac 14 F'_{\mu\nu}F'^{\mu\nu}
	+ \frac{1}{2} m^2_{A'} A'^2
	- \sum_f q_f e (A_\mu + \varepsilon A'_\mu) \bar f\gamma^\mu f
	- g_X A'_\mu \bar X\gamma^\mu X \ ,
	\label{eq:darkphoton:L}
\end{align}
where we sum over Standard Model fermions $f$ with electric charge $q_f$, $\varepsilon$ is the kinetic mixing parameter, and $g_X$ is the dark sector gauge coupling. The electromagnetic and dark fine structure constants are $\alpha = e^2/(4\pi)$ and $\alpha_X = g_X^2/(4\pi)$.

\subsection{Capture \label{SubSec:Capture}}

Dark matter is captured when it scatters off a nucleus, $N$, such that the dark matter's outgoing velocity is less than the escape velocity of Earth at the interaction point. 

\subsubsection{Theory}

The total capture rate is the sum of the individual capture rates for each element, $C^N_{\text{cap}}$,
\begin{equation}
C_{\text{cap}} = \sum_N C^N_{\text{cap}} \ .
\label{Eqn:CapRate}
\end{equation}
$C^N_{\text{cap}}$ is an integral over scattering configurations:
\begin{equation}
	C^N_{\text{cap}} = n_X
	\int_{0}^{R_{\oplus}}dr \, 4 \pi r^2 n_N(r) 
	\int_{0}^{v_{\text{gal}}} du \, 4 \pi u^2 f_{\oplus}(u) \frac{u^2 + v_{\oplus}^2(r)}{u}
	\int_{E_{\text{min}}}^{E_{\text{max}}} \, dE_R \frac{d\sigma_N}{dE_R} \Theta (\Delta E) \ .
	\label{Eqn:ReducedCapRate}
\end{equation}
The components of the integrand are as follows:
\begin{itemize}
	\item $E_R$ is the recoil energy transferred from the incident dark matter onto the target nucleus in the frame of the Earth.

	\item $\Delta E = E_\text{max} - E_\text{min}$ is the difference between the maximum kinematically allowed recoil energy and the minimum recoil energy required for dark matter to capture, (\ref{Eqn:Eminmax}). $\Theta(\Delta E)$ restricts the integrand to kinematically allowed scattering processes where the dark matter is captured in the Earth.
	
	\item ${d\sigma_N}/{dE_R}$ is the cross section for dark matter to scatter off of a nucleus $N$. We assume elastic scattering so that in the non-relativistic limit~\cite{Feng:2015hja},
	\begin{equation}
		\frac{d\sigma_N}{dE_R} \approx 
	8 \pi \epsilon^2 \alpha_X \alpha Z_N^2 \frac{1}{(u^2 + v_{\oplus}^2)} \frac{m_N}{(2m_NE_R + m_{A'}^2)^2} \left | F_N(E_R) \right|^2 \ .
	\label{Eqn:CrossSection}
	\end{equation}
	$F_N(E_R)$ is the Helm form factor, encoding the coherent scattering of the dark matter off multiple nucleons in the nucleus, $N$. We parameterize this as $\left| F_N(E_R) \right| ^2 = e^{-E_R/E_N}$, where $E_N \equiv 0.114~\text{GeV}/A_N^{5/3}$ for a nucleus with atomic mass number $A_N$~\cite{Lewin:1995rx}. 
	
	\item $v_{\oplus}(r)$ is the escape velocity of Earth at radius $r$. 
	
	\item $f_{\oplus}(u)$ is the dark matter velocity distribution in the Earth's reference frame.
	
	\item $n_N(r)$ is the density of element $N$ in the Earth at radius $r$. 
	
	\item $n_X = (0.3/\text{cm}^3)(\text{GeV}/m_X)$ is the local dark matter density in the Milky Way \cite{2010A&A...509A..25W}.
\end{itemize}
The integrations in (\ref{Eqn:ReducedCapRate}) run over the radial distance from the center of the Earth $r$, the dark matter velocity $u$ in the Earth frame, and the recoil energy $E_R$ imparted by the dark matter to the target nucleus $N$.

The lower limit of the recoil energy integral $dE_R$ is the minimum energy transfer required for the dark matter particle of velocity $u$ to capture; this is the kinetic energy of the free dark matter particle. The upper bound is the maximum kinematically accessible recoil energy allowed by elastic scattering in the frame of the target nucleus. These energies are:
\begin{align}
E_{\text{min}}(u) &= \frac{1}{2}m_X u^2
&
E_{\text{max}}(u, r) &= \frac{2\mu_N^2}{m_N} \left(u^2 + v_{\oplus}^2(r)\right) \ .
\label{Eqn:Eminmax}
\end{align}

The $du$ integral over dark matter velocities in the Earth frame is weighted by the angular-and time-averaged dark matter velocity distribution~\cite{Vergados:1998ax}:
\begin{equation}
f_\oplus(u) 
= 
\frac{1}{4} 
\int_{-1}^{1} 
dc_\theta
\int_{-1}^{1}
dc_\phi 
\ f\left[ 
	(u^2 + 
		(V_\odot + V_\oplus c_\gamma c_\phi)^2 + 
		2u(V_\odot + V_\oplus c_\gamma c_\phi)
	c_\theta)^{1/2} 
	\right]\ ,
\label{Eqn:fcross}
\end{equation}
where we write $c_\varphi \equiv \cos(\varphi)$. 
The velocity of the Sun relative to the galactic center is $V_\odot\approxeq 220$ km/s and the velocity of the Earth relative to the Sun is $V_\oplus \approxeq 29.8$ km/s. The angle $\gamma$ in $c_\gamma\approx 0.51$ is the inclination between the orbital planes of the Earth and Sun. Numerical simulations and   observations suggest a velocity distribution~\cite{Vogelsberger:2008qb, Fairbairn:2008gz, Kuhlen:2009vh, Ling:2009eh, Lisanti:2010qx, Mao:2012hf}: 
\begin{equation}
f(u) = N_0 \left[ \exp \left( \frac{v_\text{gal}^2 -u^2}{ku_0^2} \right) -1 \right]^k \Theta(v_\text{gal} - u)
\label{Eqn:f0}
\end{equation}
where $N_0$ is the normalization, $v_\text{gal}$ is the escape velocity from the galaxy, and the parameters describing of distribution have typical values in the ranges~\cite{Baratella:2013fya}:
\begin{equation}
220 \ \text{km/s} < u_0 < 270 \ \text{km/s} \qquad
450 \ \text{km/s} < v_\text{gal} < 650 \ \text{km/s} \qquad
1.5 < k < 3.5
\end{equation}
The $dr$ integral accounts for the scattering location which may occur anywhere inside the volume of the Earth.

\subsubsection{Implementation}

The goal of this work is to produce plots of signal events (pairs of leptons coming from mediator decay) in $(m_{A'},\varepsilon)$ space. This requires numerically evaluating the capture rate, \cref{Eqn:ReducedCapRate}, at many values of $m_{A'}$, which is computationally expensive. 
We now demonstrate a that appropriate limits may be taken to reduce the number of integrations to greatly improve the computational efficiency of this calculation. 

\begin{figure}
	\centering
	\subfloat[Integration region for $0\leq u \leq v_{\text{gal}}$]{\includegraphics[width=0.47\textwidth]{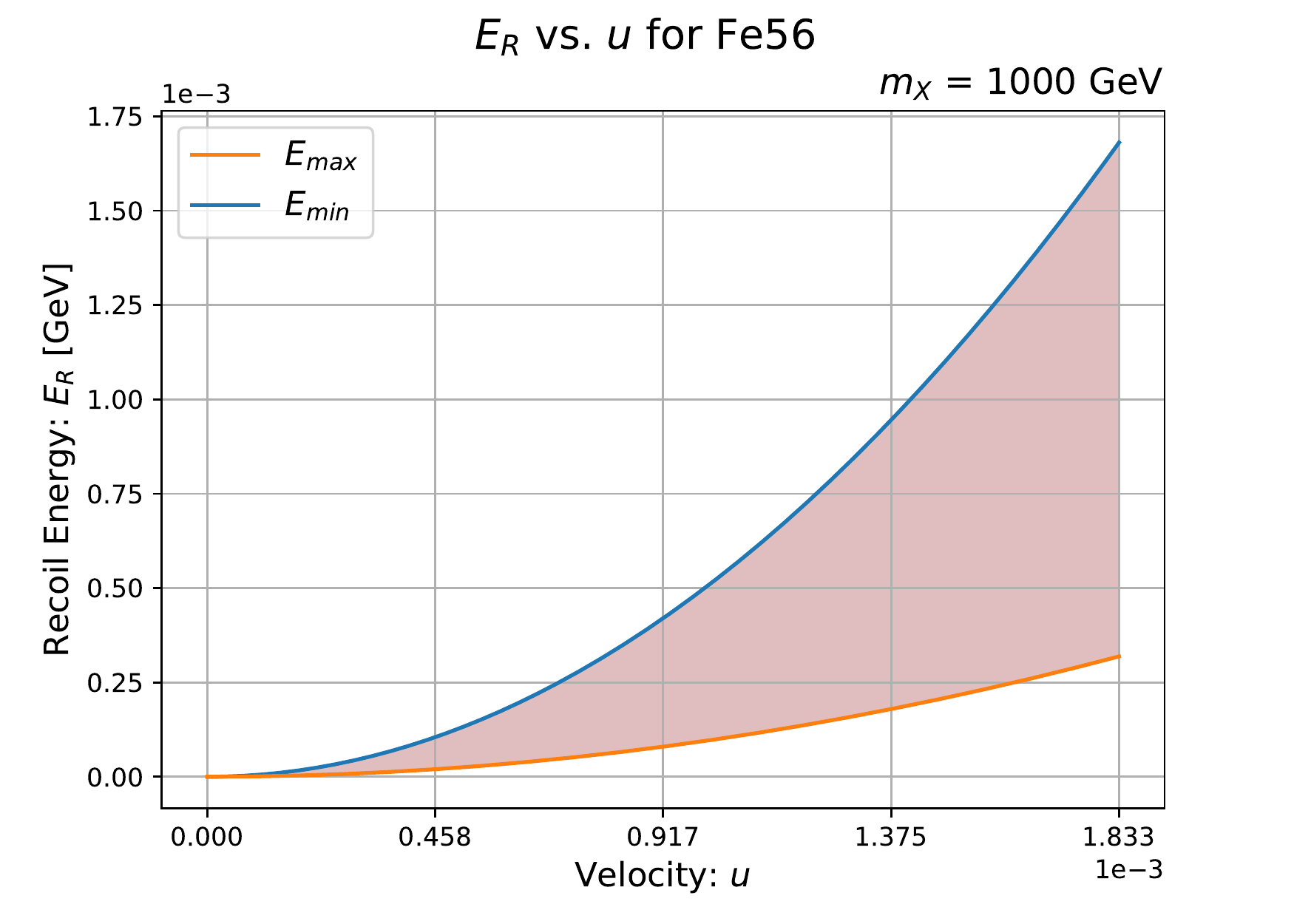} \label{Fig:recoilA}}
	\quad
	\subfloat[Integration region for $0\leq u \leq u_\text{int}$]{\includegraphics[width=0.47\textwidth]{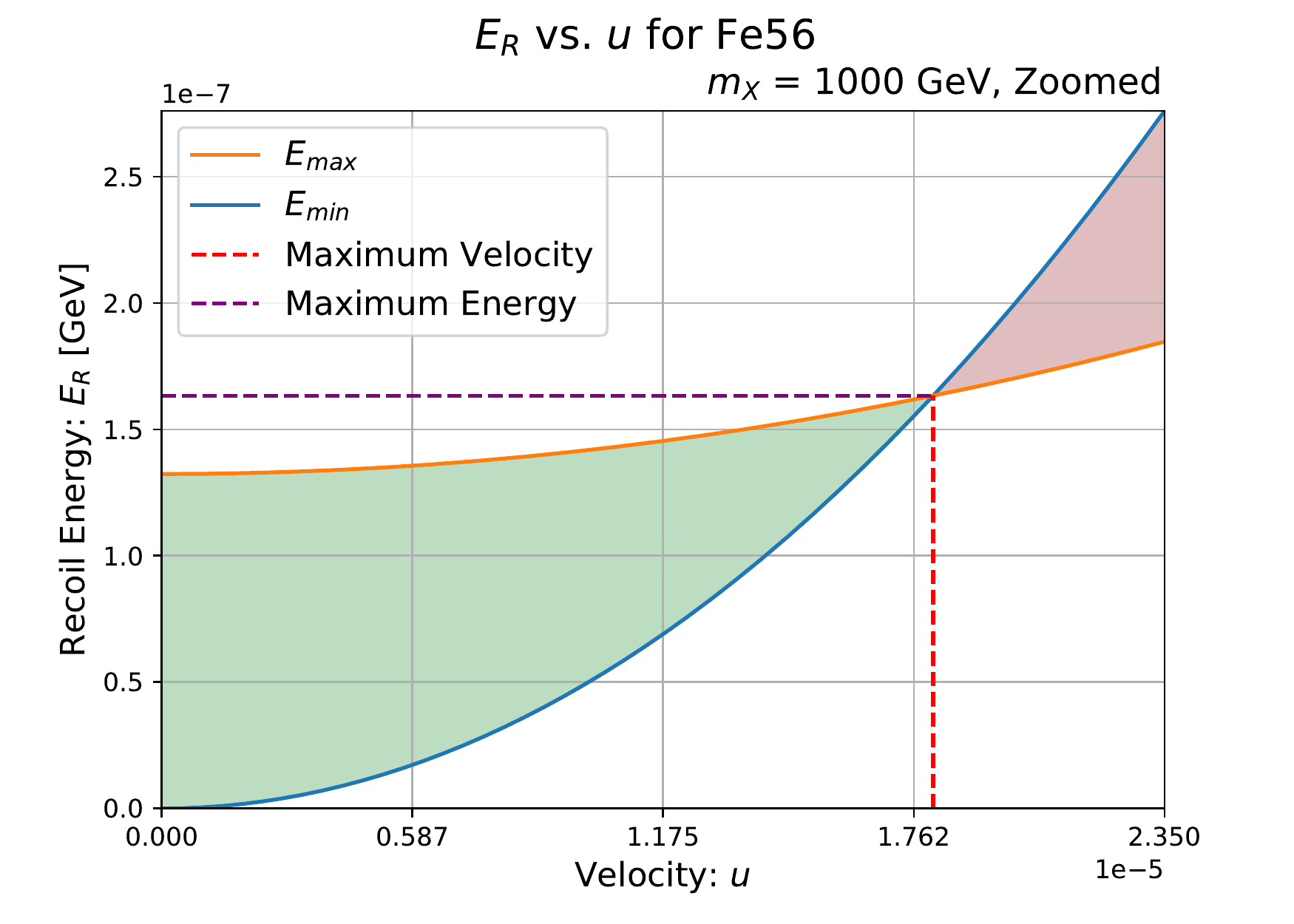}\label{Fig:recoilB}}
	\caption{Plots displaying the integration area for the velocity and recoil energy integrals. 
	The region shaded in red represents velocities where capture is kinematically forbidden and the region shaded in green represents velocities where capture is allowed.
	For nearly the entire integral, $E_{\text{min}}(u) > E_{\text{max}}(u)$ and the integral over recoil energy vanishes. However, there is a very small region near the origin for which the integral is non-vanishing.
	We take $v_{\oplus}^2 = 1.40\times 10^{-9}$,  the average escape velocity of Earth, and $m_N = 56$~GeV for the mass of the target Iron nucleus. 
	}
	\label{Fig:recoilEnergies}
\end{figure}

First, observe that the upper limit of the $dE_R$ integral in \cref{Eqn:ReducedCapRate} is actually much larger than the region for which the integrand is non-zero. This is because for most of the integration region, $E_\text{min}(u) > E_\text{max}(u)$, where we recall from \cref{Eqn:Eminmax} that $E_\text{min}$ comes from the requirement of capture and $E_\text{max}$ comes from kinematics. This means that the $\Theta(\Delta E)$ factor forces the integrand to vanish for most of the integral, except for a small region at low $E_R$. Mathematically, both $E_\text{min}(u)$ and $E_\text{max}(u)$ are parabolas, but $E_\text{max}(u)$ has a shallower curvature and a small positive constant shift relative to $E_\text{min}(u)$; see \cref{Fig:recoilEnergies}. Thus it is sufficient to only integrate over a much smaller range of recoil energies $E_R$ and velocities $u$. 

The velocity, $u_\text{int}(r)$, where $E_\text{max}(u_\text{int}) = E_\text{min}(u_\text{int})$, represents the fastest velocity a dark matter particle may have and still be captured at radius $r$, 
\begin{align}
u_{\text{Int}} &= \sqrt{ \frac{B}{A-B}}\,v_{\oplus}(r)	
&
A&=\frac{1}{2}m_X
&
B&=2 \frac{\mu_N^2}{m_N} \ .
\label{Eq:Eminmax:matching}
\end{align}
Any dark matter particle with velocity larger than $u_\text{int}$ will have an outgoing velocity greater than Earth's escape velocity and it will not be captured.  

This simplification is demonstrated in \cref{Fig:recoilEnergies}, where we plot recoil energy against incident dark matter velocity. In \cref{Fig:recoilB}, we see that the largest typical velocity a dark matter particle may have and still be captured is $\mathcal{O}(10^{-5})$ in natural units. Therefore, integrating to $v_\text{gal} \approx 1.8 \times 10^{-3}$ is computationally wasteful. The $\Theta(\Delta E_R)$ that imposes a strict cutoff on the region of non-zero integrand can be difficult for standard numerical integration methods, especially when large portions of the integration region are zero.

A second way to boost the efficiency of the integration in \cref{Eqn:ReducedCapRate} is to observe that the $E_R$ term in the denominator of \cref{Eqn:CrossSection} is always subdominant to the $m_{A'}^2$ term. The largest momentum transfer occurs for Ni$^{58}$ and is $2m_NE_R \sim \mathcal{O}(10^{-5} \ \text{GeV}^2)$. This is smaller than the lightest dark photon mass we consider, $m_{A'}^2 \sim \mathcal{O}(10^{-4} \ \text{GeV}^2)$. We can thus approximate the scatter as a point-like interaction, where the momentum dependence (form factor) of the cross section is negligible, $2m_NE_R \ll m_{A'}^2$. This lets us write the propagator as:
\begin{equation}
\frac{1}{(2m_NE_R + m_{A'}^2) } \rightarrow \frac{1}{m_{A'}^2}
	\label{eq:lowE:approx}
\end{equation}
where we model dark matter scattering off of a nucleus by an effective interaction with a coupling proportional to $1/m_{A'}^2$. Critically for the computational efficiency of this calculation, this removes the non-trivial $E_R$ dependence of the integrand in \cref{Eqn:ReducedCapRate}.

With this approximation, we able to write the capture rate as the product of two functions; one of which depends only on $(m_{A'}, \varepsilon, \alpha_X)$ and the other which only depends on $(m_X, \alpha)$:
\begin{equation}
C_{\text{cap}} = \left( \frac{\varepsilon^2}{m_A^4} \alpha_X \right) \kappa_0(m_X,\alpha) \ .
\label{Eqn:CapRateKappa}
\end{equation}
The quantity $\kappa_0$ is simply a regrouping of \cref{Eqn:ReducedCapRate} after imposing the small recoil energy approximation, \cref{eq:lowE:approx}:
\begin{equation}
\kappa_0(m_X, \alpha) \equiv \sum_N \left[ n_X
\int_{0}^{R_{\oplus}}dr \, 4 \pi r^2 n_N(r) 
\int_{0}^{u_{\text{Int}}} du \, 4 \pi u^2 f_{\oplus}(u) \frac{u^2 + v_{\oplus}^2(r)}{u}
\int_{E_\text{min}}^{E_\text{max}} \, dE_R \frac{d\varsigma_N}{dE_R} \right]
\label{Eqn:Kappa0}
\end{equation}
with 
\begin{equation}
\frac{d\varsigma_N}{dE_R} = 8 \pi \alpha Z_N^2 \frac{1}{(u^2 + v_{\oplus}^2)} m_N \left | F_N \right|^2 \ .
\end{equation}
In this form, the quantity $\kappa_0$ is a constant in $(m_{A'}, \varepsilon)$ space and only needs to be calculated once for a given dark matter mass $m_X$. Consequently, for a single point, the computation time between \cref{Eqn:ReducedCapRate} and \cref{Eqn:CapRateKappa} is comparable. However, using 10 test points, the computation time using the $\kappa_0$ formulation is an order of magnitude faster.

We use density data from the Preliminary Earth Reference Model\footnote{The original data file is available at \url{http://ds.iris.edu/spud/earthmodel/9991844}.}~\cite{PREM500Paper} for the dominant elements that compromise the Earth at a discrete set of 491 radii. As such, we replace the $dr$ integral over the radius of the Earth with a discretized sum.

\subsection{Annihilation}

After becoming gravitationally captured, dark matter re-scatters off matter in the Earth, losing kinetic energy until it falls to the center of the Earth where it thermalizes with the surrounding matter.

\subsubsection{Theory}

The annihilation rate for thermalized dark matter is \cite{Baratella:2013fya}:
\begin{equation}
	C_{\text{ann}} = \langle \sigma_{\text{ann}} v \rangle \left[ \frac{G_N m_X \rho_{\oplus}}{3T_{\oplus}} \right]^{3/2} \ ,
	\label{Eqn:AnnRate}
\end{equation}
where $\rho_{\oplus}\approx 13 \ \text{g}/\text{cm}^3$ is the density at the center of the Earth, and $T_\oplus \approx 5700$~K is the temperature at the center of the Earth. $\sigma_\text{ann}$ is the cross section for $XX \rightarrow A'A'$ and $v$ is the relative velocity between the two dark matter particles.
The thermally-averaged cross section is:
\begin{equation}
	\langle \sigma_{\text{ann}} v \rangle = (\sigma_{\text{ann}} v)_\text{tree} \langle S_S \rangle
	\label{Eqn:SigmaVAvg}
\end{equation}
where the tree level cross section is given by \cite{Liu:2014cma}
\begin{equation}
	(\sigma_{\text{ann}}v)_\text{tree} = \frac{\pi \alpha_X^2}{m_X^2} \frac{\left[ 1- m_{A'}^2/m_X^2 \right]^{3/2}}{ \left[ 1- m_{A'}^2/(2m_X^2) \right]^2 } \ .
	\label{Eqn:SigmaVTree}
\end{equation}
The dark sector fine structure constant, $\alpha_X$, is fixed by requiring that dark matter is a thermal relic: $\langle\sigma_\text{ann} v\rangle = 2.2 \times 10^{-26}~\text{cm}^3/\text{s}$~\cite{Steigman:1997vs}.
$\langle S_S \rangle$ is the thermally averaged $s$-wave Sommerfeld enhancement. The analytic expression for this enhancement is found by approximating the Yukawa potential for the Hulth\'en potential \cite{Cassel:2009wt,Feng:2010zp}. The resulting Sommerfeld factor is:
\begin{equation}
	S_S = \frac{\pi}{a} \frac{\sinh(2\pi ac)}{\cosh(2\pi ac) - \cos(2\pi \sqrt{c-a^2c^2})}
	\label{Eqn:Somm}
\end{equation} 
where $a=v/2\alpha_X$ and $c = 6\alpha_X m_X / (\pi^2 m_{A'})$. The thermal average is then \cite{Feng:2015hja}:
\begin{equation}
	\langle S_S \rangle = \int \frac{d^3v}{(2\pi v_0^2)^{3/2}} e^{-\frac{1}{2}v^2/v_0^2} S_S
	\label{Eqn:SommAvg}
\end{equation}
with $v_0 = \sqrt{2T_\oplus/m_X}$.

\begin{figure}[H]
	\centering
	\subfloat{\includegraphics[width=0.4\textwidth]{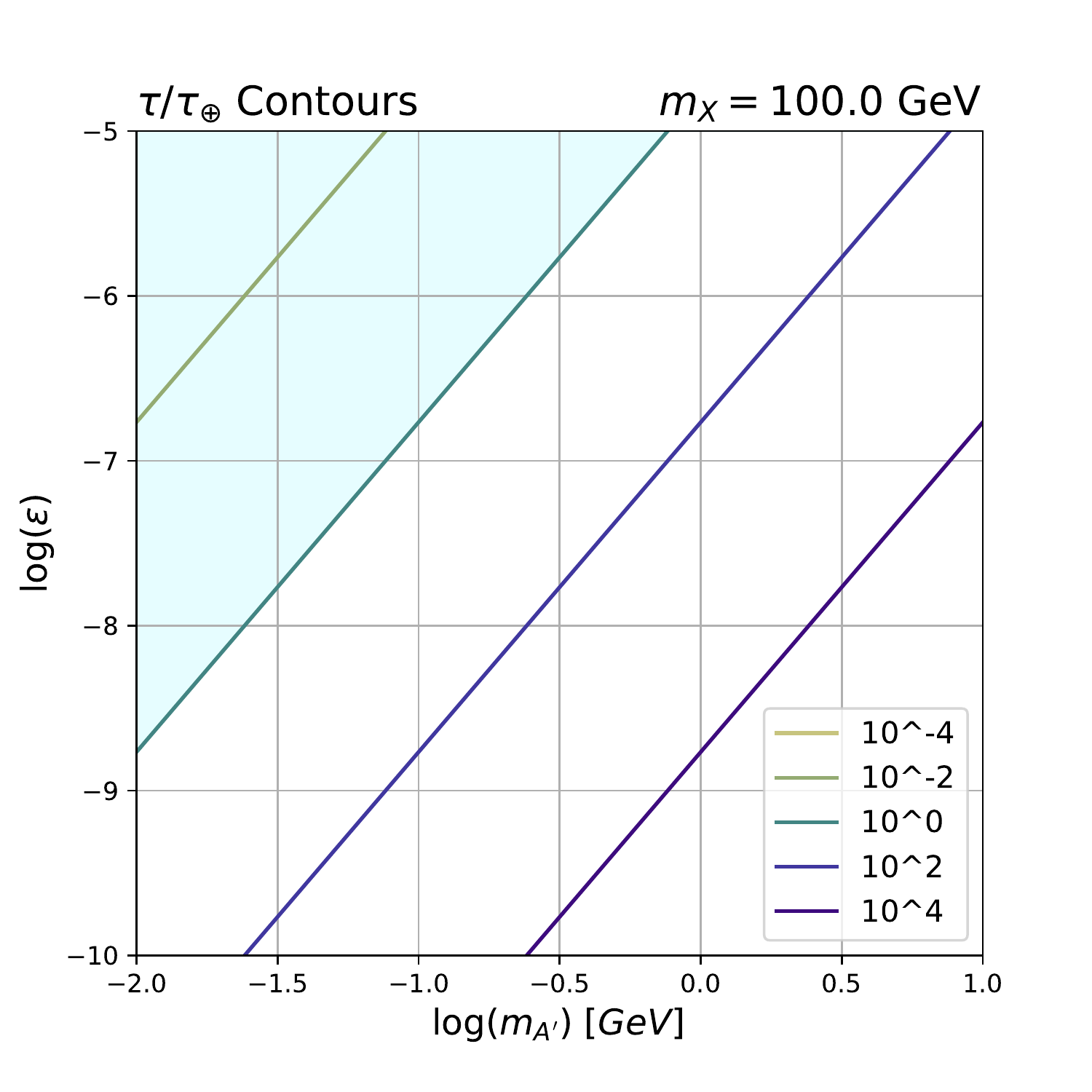}}
	\quad
	\subfloat{\includegraphics[width=0.4\textwidth]{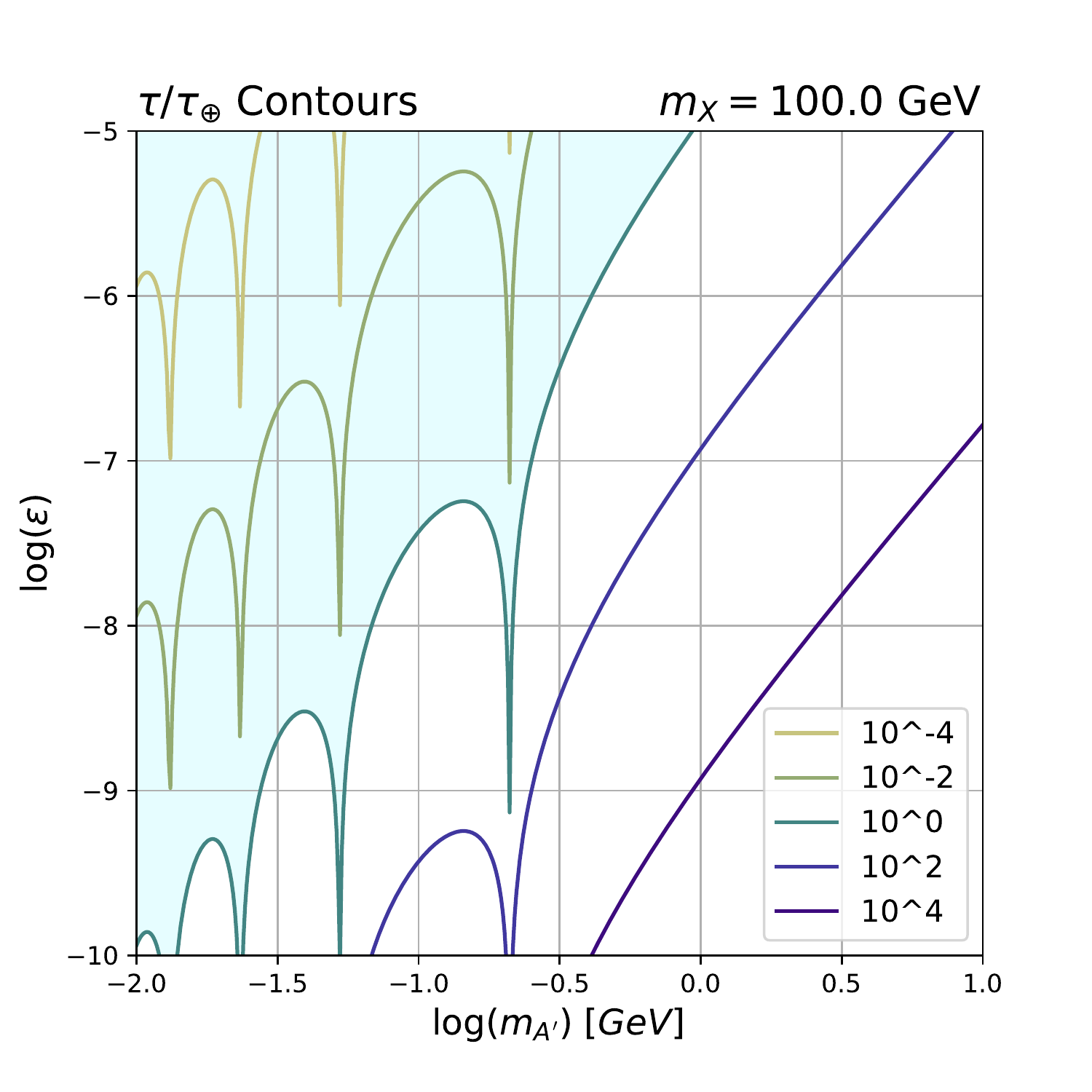}}
	\\
	\subfloat{\includegraphics[width=0.4\textwidth]{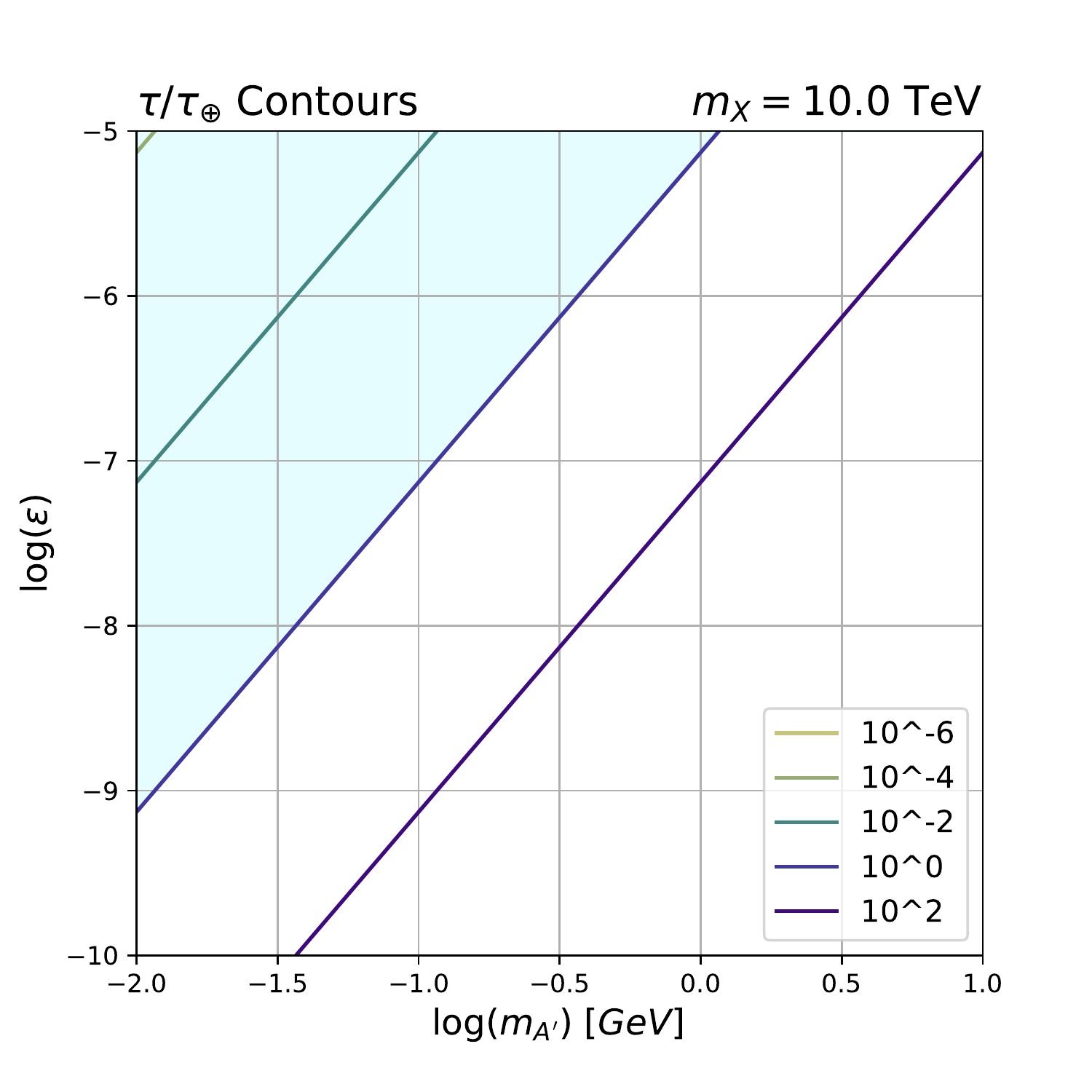}}
	\quad
	\subfloat{\includegraphics[width=0.4\textwidth]{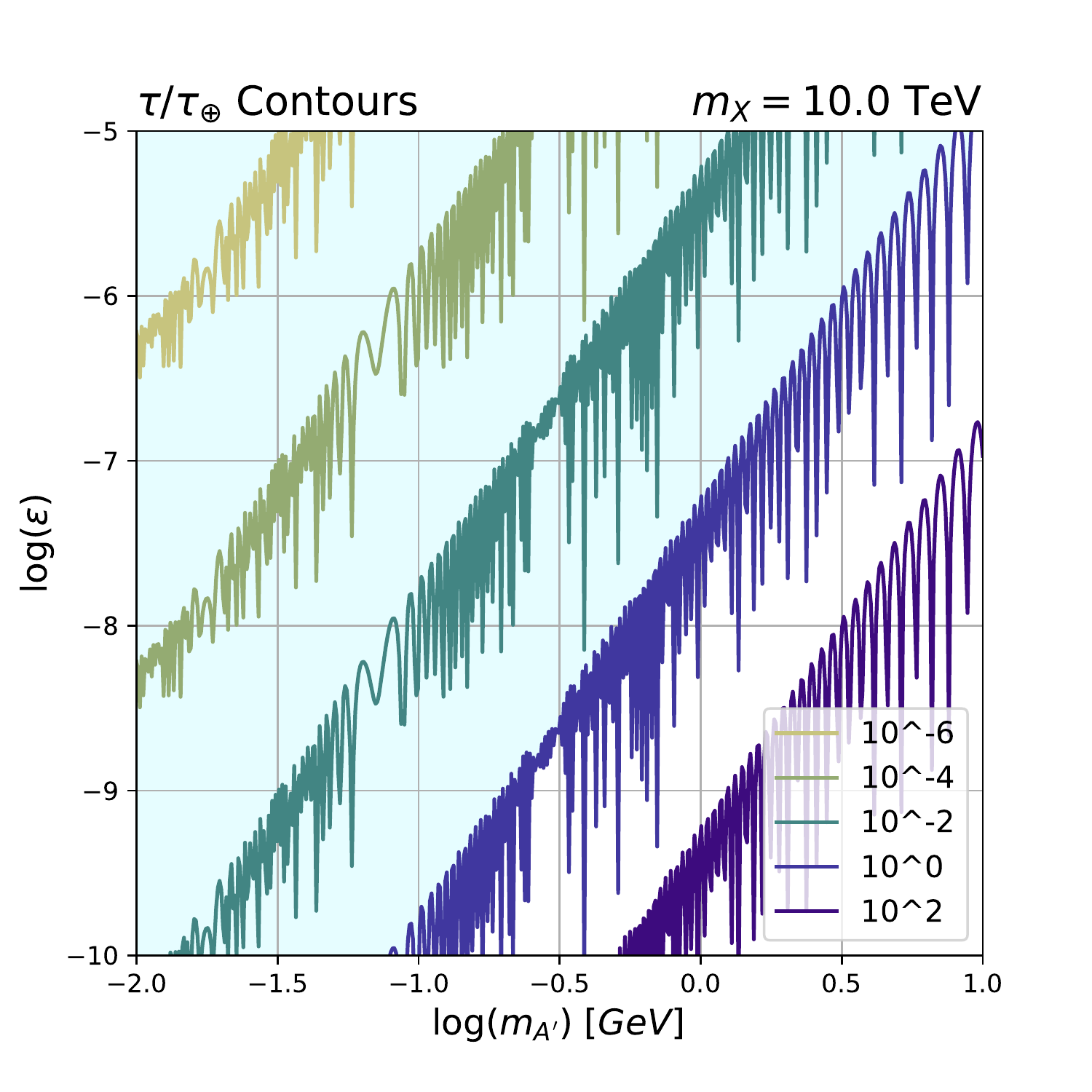}}
	\caption{Plots of dark matter masses, $m_X = 10 \ \text{GeV}$ and $m_X = 10 \ \text{TeV}$ with and without Sommerfeld effects. We can see that as the dark matter mass increases, the Sommerfeld enhancement becomes more prominent, pushing the contours towards smaller $\varepsilon$ and larger $m_{A'}$. These plots reproduce the results in~\cite{Feng:2015hja}.}
	\label{Fig:contourPlots}
\end{figure}

\subsubsection{Implementation}

Because the Sommerfeld enhancement is independent of the kinetic mixing, it is computationally frugal to calculate the Sommerfeld enhancement for a given dark matter mass---and hence a given $\alpha_X$ from the thermal relic condition---as a function of the mediator mass. This data can then be saved separately to be called for parameter scans in the $(m_{A'}, \varepsilon)$ plane. 

\subsection{Equilibrium Timescale}
\label{Sec:EqTimescale}

The equilibrium time,
\begin{align}
\tau &\equiv \frac{1}{\sqrt{C_{\text{cap}}C_{\text{ann}}} } \ ,
\label{Eq:equilibrium:time}
\end{align}
is the characteristic amount of time it takes the dark matter population within Earth to approach its maximal value. 
The relevant timescale for comparison is the age of the Earth. If the age of the Earth is at least $\tau$, then dark matter capture and annihilation has equilibrated and the rate at which captured dark matter annihilates into mediators is maximized.

\subsubsection{Theory}

In \cref{Fig:contourPlots} we present plots of constant $\tau/\tau_{\oplus}$ in the $(m_{A'}, \varepsilon)$ plane for $m_X = 100 \ \text{GeV}$ and $10 \ \text{TeV}$. The $m_{A'}$ and $\varepsilon$ dependence enters $\tau$ primarily through the capture rate, $\tau \sim C_\text{cap}^{-1/2} \sim m_{A'}^2/\varepsilon$. The plots on the left represent contours without Sommerfeld effects. Once the Sommerfeld effect is included, resonant structures appear, shown in the plots on the right. The shaded blue region is the region of parameter space where the population of dark matter in the Earth has reached its maximal and equilibrium value. Most notably, the Sommerfeld enhancements pushe these contours farther into parameter space by factors of $10-10^4$ for $m_X \sim 100 \ \text{GeV} - \ 10 \ \text{TeV}$.

\subsubsection{Implementation}

The small recoil energy approximation also makes the computation of the equilibrium time more efficient. To see this, we start with a contour level of $\tau/\tau_{\oplus} = 10^\mathrm{L}$. Writing the capture rate as \cref{Eqn:CapRateKappa} and the annihilation rate as $C_\text{ann} = C_\text{ann,0} \langle S_S \rangle$, with $C{_{\text{ann,0}}}$ being the annihilation rate without any Sommerfeld enhancements, we are able to derive the following expression for the kinetic mixing parameter as a function of mediator mass:
\begin{equation}
	\log(\varepsilon) = 2\log(m_{A'}) -\frac{1}{2}\log(\alpha_X C_\text{ann,0} \langle S_s \rangle) - {L} - \frac{1}{2}\log(\kappa_0) - \log(\tau_{\oplus})
	\label{Eqn:EpsilonFunction}
\end{equation}
The first two terms in \cref{Eqn:EpsilonFunction} carry all of the $m_{A'}$ dependence, ${L}$ corresponds to the contour level of $10^{L}$ and the last two terms are constants in $(m_{A'}, \varepsilon)$ space. In this form, each contour of constant $\tau/\tau_{\oplus}$ is encoded in a function $\varepsilon(m_{A'})$, offset by the contour level $10^L$.

\subsection{Dark Photon Signal Characteristics}

Boosted mediators produced in dark matter annihilation propagate outward through the capturing body. These may then decay near the surface into collimated pairs of light, Standard Model particles. These particles may leave upward-going tracks in large-volume detectors. \texttt{DarkCapPy} assumes a dark photon mediator that decays to pairs of leptons. For a given dark matter mass $m_X$, mediator mass $m_{A'}$, and mixing parameter $\varepsilon$, the output is a number of mediator decays inside the volume of the detector (IceCube) over the observation time, $T$.

\subsubsection{Theory}
	
Dark photons decay to Standard Model fermions with width \cite{Buschmann:2015awa}:
\begin{equation}
	\Gamma(A'\rightarrow f \bar{f}) = \frac{N_C \varepsilon^2 q_f^2 \alpha (m_{A'}^2 + 2m_f^2)}{3m_{A'}} \sqrt{1-\frac{4m_f^2}{m_{A'}^2}}
	\label{Eqn:DecayWidth}
\end{equation}
where $N_C$ are the number of colors of fermion $f$. For $m_{A'} \gg m_e$, the boosted dark photons decay to $f\bar f$ pairs with characteristic length:
\begin{equation}
	L = R_{\oplus}B_e \left( \frac{3.6\times 10^{-9}}{\varepsilon} \right)^2 \left( \frac{m_X/m_{A'}}{1000} \right) \left( \frac{\text{GeV}}{m_{A'}} \right)
	\label{Eqn:DecayLength}
\end{equation}
where $R_{\oplus} \approxeq 6370$ km is the radius of Earth, $B_e \equiv B(A' \rightarrow e^+e^-)$ is the dark photon branching fraction to electrons and the quantity $m_X/m_{A'}$ is the Lorentz factor resulting from a decay of a heavy initial state into a light final state. For $m_X \sim \text{GeV} - \text{TeV}$, and $m_{A'} \sim 100 \ \text{MeV} - \text{GeV}$, the requirement that $L\sim R_{\oplus}$ yields $\varepsilon \sim 10^{-11} - 10^{-7} $. We expect the largest number of signal events in this region of parameter space.

\begin{figure}[H]
	\centering
	\subfloat[Without Sommerfeld effects.]{\includegraphics[width = 0.4\textwidth]{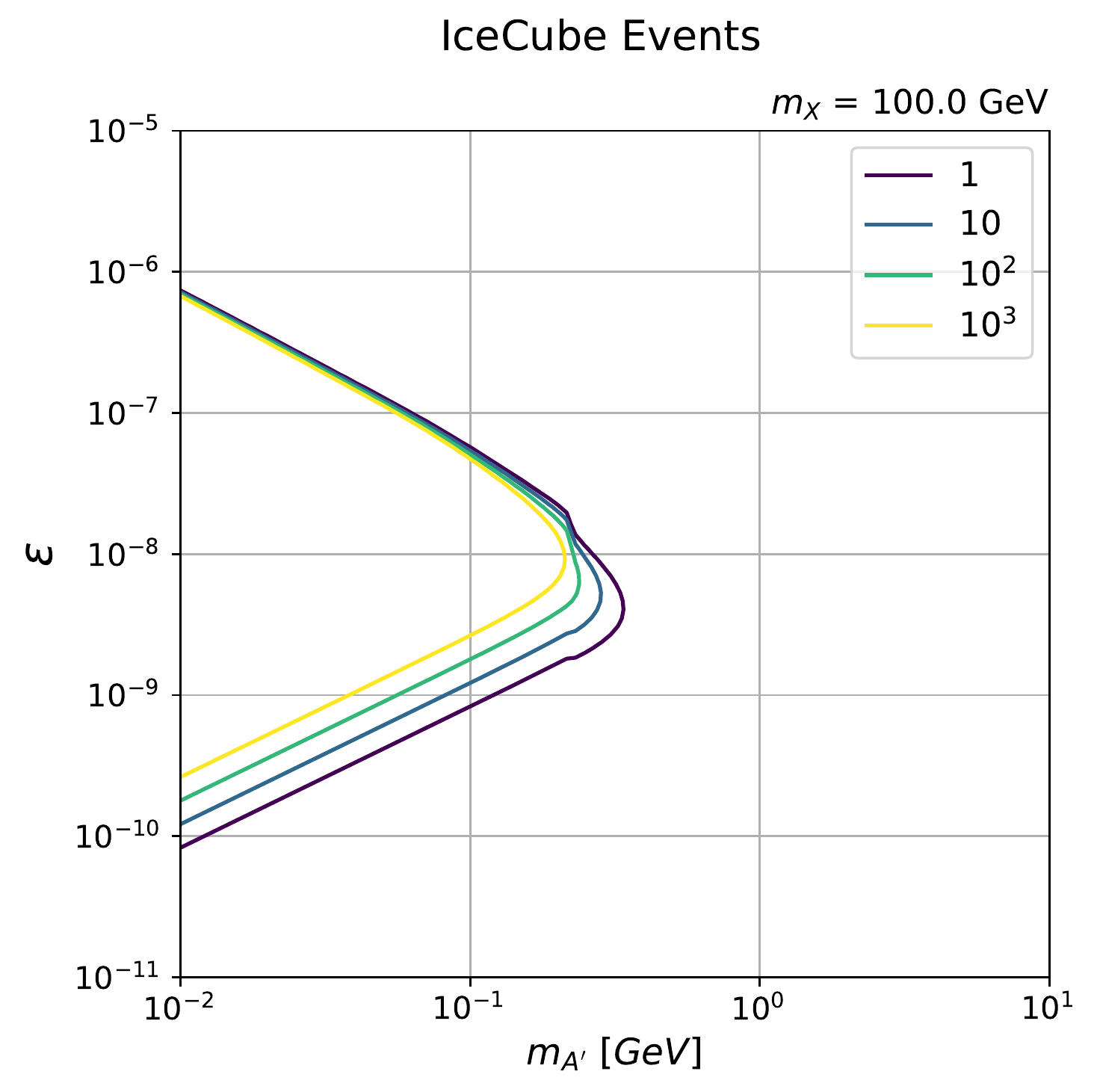}}
	\qquad
	\subfloat[With Sommerfeld effects.]{\includegraphics[width = 0.4\textwidth]{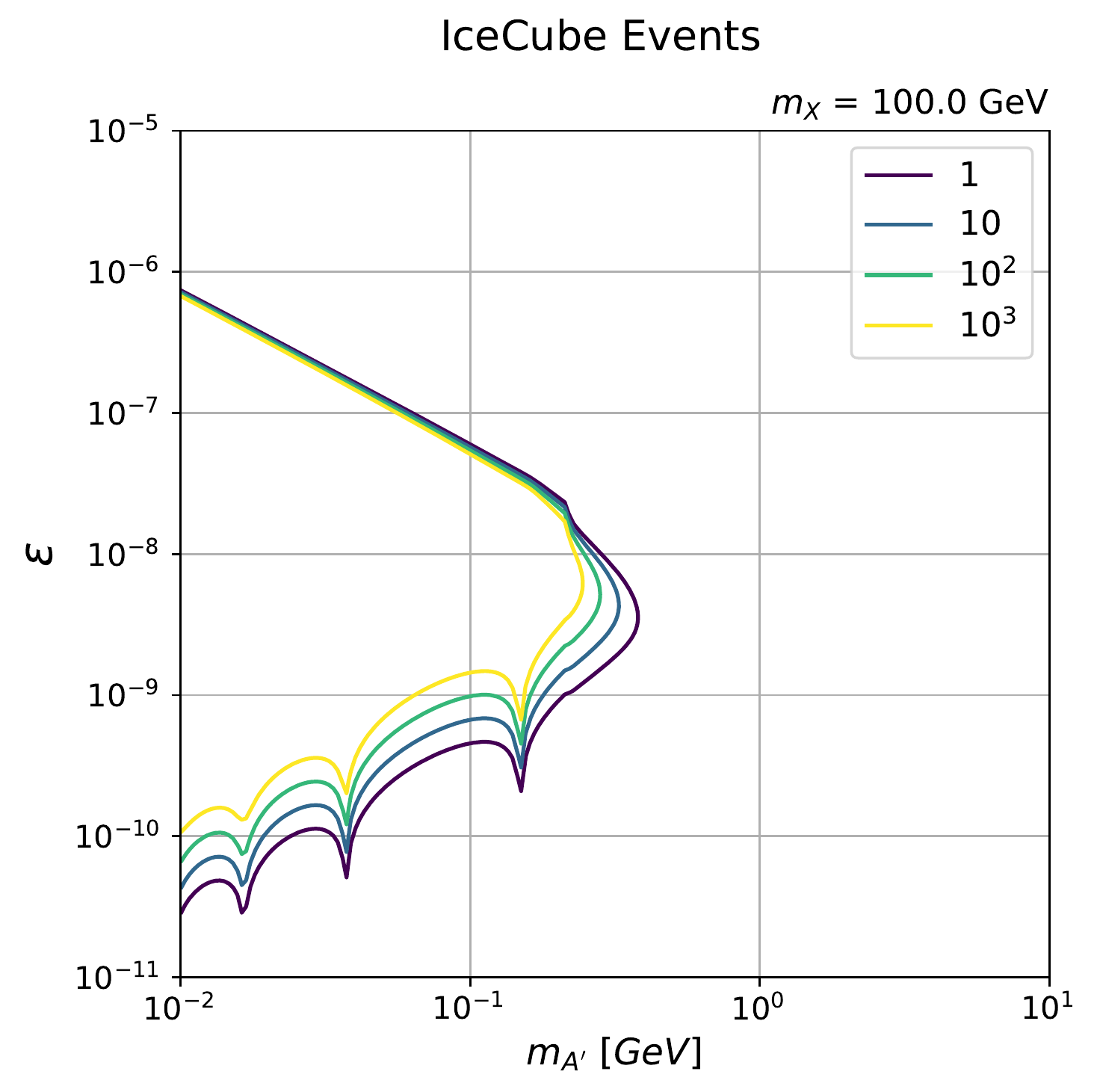}}
	\caption{Number of signal events at IceCube for an observation time of $T = 10$ years.}
	\label{Fig:IceCubeSignal}
\end{figure}
The number of dark photon decays that can be detected is \cite{Feng:2015hja}:
\begin{equation}
	N_{\text{sig}} = 2 \Gamma_\text{ann} \frac{A_\text{eff}}{4\pi R_{\oplus}^2} \epsilon_{\text{decay}}T \ ,
	\label{Eqn:Nsig}
\end{equation}
where $A_\text{eff} = 1$ km$^2$ is the effective area of the detector~\cite{Ahrens:2002dv}. We set the default observation time of the detector to be $T = 10$ years. The decay parameter \cite{Feng:2015hja},
\begin{equation}
	\varepsilon_\text{decay} = e^{-R_\oplus/L} - e^{-(R_\oplus + D)/L} \ ,
	\label{Eqn:EpsilonDecay}
\end{equation}
is the probability that a dark photon decays between a distance $R_\oplus$, the radius of Earth, and $R_\oplus + D$, where $D$ is the depth of the detector. For IceCube, we take $D \approx 1$ km~\cite{Ahrens:2002dv}.

\subsubsection{Implementation}

Because the relative couplings to Standard Model fermions is fixed by the fermion electric charges, the branching ratio of a dark photon to a given fermion species depends only on the dark photon mass. This dependence, however, can become complicated when decays to hadrons are kinematically accessible. We extract the relevant branching ratio from \cite{Buschmann:2015awa} as a csv file that is input into \texttt{DarkCapPy}. 

In \cref{Fig:IceCubeSignal}, we present plots of the number of detectable dark photon decays at IceCube with a 10 year observation time. The plot on the left does not contain Sommerfeld effects, while the plot on the right does. The Sommerfeld effects push the contours roughly one order of magnitude towards smaller $\varepsilon$. When the mixing parameter is large, the contours are primarily fixed by the characteristic decay length, $L\sim 1/(\varepsilon^2 m_{A'}^2)$. When the mixing parameter is small, the lower contours are determined by the timescale it takes $\tanh^2(\tau_{\oplus}/\tau)$ to reach unity. These results match \cite{Feng:2015hja}.

\section{Installing}

This package is hosted on GitHub\footnote{\url{https://github.com/agree019/DarkCapPy}}; all program files may be accessed there. The most convenient way to install this package is through the \texttt{pip} Python package management system. To do this, run the following \texttt{pip} command at command line:
\begin{center}
	\texttt{pip install git+https://github.com/agree019/DarkCapPy}
\end{center}
This installs \texttt{DarkCapPy} in your Python path. 

\section{Package Contents}

\begin{figure}[H]
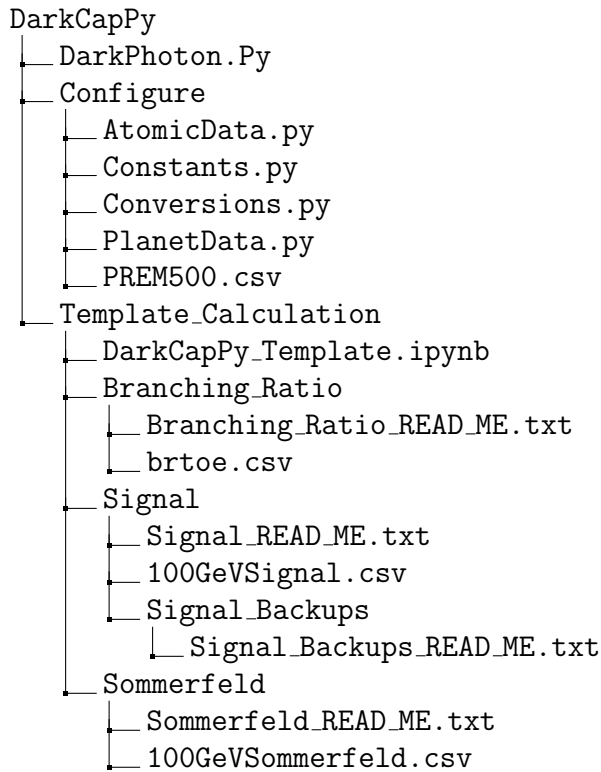

	\dirtree{%
		.1 DarkCapPy.
		.2 DarkPhoton.Py.
		.2 Configure.
		.3 AtomicData.py.
		.3 Constants.py.
		.3 Conversions.py.
		.3 PlanetData.py.
		.3 PREM500.csv.
		.2 Template\_Calculation.
		.3 DarkCapPy\_Template.ipynb.
		.3 Branching\_Ratio.
		.4 Branching\_Ratio\_READ\_ME.txt.
		.4 brtoe.csv.
		.3 Signal.
		.4 Signal\_READ\_ME.txt.
		.4 100GeVSignal.csv.
		.4 Signal\_Backups.
		.5 Signal\_Backups\_READ\_ME.txt.
		.3 Sommerfeld.
		.4 Sommerfeld\_READ\_ME.txt.
		.4 100GeVSommerfeld.csv.
	}
	\caption{Directory structure of \texttt{DarkCapPy} package.}
	\label{fig:DirectoryTree}
\end{figure}

\cref{fig:DirectoryTree} outlines the directory structure of the package. We provide a Jupyter notebook that implements this package in the folder \texttt{Template\_Calculation}.

\setlength{\DTbaselineskip}{14pt}

\section{Functions}

We briefly describe what each function does, how it is defined, and how it is called.
 We provide Python function definitions typeset in \texttt{teletype font}.
We list function definitions in order of dependence, so later functions depend only on those defined earlier in this document.

\subsection*{\texttt{DarkPhoton.py}}

This file contains all the mathematical definitions to calculate the Earth capture scenario via a massive dark photon \cite{Feng:2015hja}. 
Prefactors that are removed from the intermediate computations for computational efficiency are restored at the end of the calculation. In this section, we account for these prefactors by listing them under \emph{factors removed}. 

\subsubsection*{Capture Functions}
\begin{itemize}
	\item \textbf{Helm Form Factor}: The Helm form factor accounts for the coherent scattering off multiple nucleons in the target nucleus. A dark matter particle scattering off a nucleus $N$ with recoil energy $E_R$ has a Helm form factor,
	\begin{equation}
		\texttt{formFactor2}(N,E_R) = \left| F_N(E_R) \right| ^2 = e^{-E_R/E_N}
		\label{eqn:formFactor}
	\end{equation}
	where $E_N = 0.114 \ \text{GeV}/A_N^{3/5}$ is the characteristic energy scale for an atom with atomic mass $A_N$.
	
	\item \textbf{Cross Section}: Defines the differential cross section for elastic $XN\rightarrow XN$ scattering as a function of recoil energy, \cref{Eqn:CrossSection}. 
	
	\begin{equation}
		\texttt{crossSection}(N, m_{A'},E_R) = \frac{m_N}{(2m_N E_R + m_{A'}^2)^2} \left| F_N(E_R) \right|^2
		\label{eqn:crossSection}
	\end{equation}
	
	\emph{Factors Removed}: $8\pi \alpha_X \alpha \epsilon^2 Z_N^2$.  The part used to calculate $\kappa_0$, \cref{Eqn:Kappa0}, is:
	\begin{equation}
		\texttt{crossSetionKappa0}(N, E_R) = \left| F_N(E_R) \right| ^2
		\label{eqn:crossSectionKappa0}
	\end{equation}
	\emph{Factors Removed}: $(8\pi \alpha_X \alpha m_N Z_N^2){\varepsilon^2 \alpha_X}/{m_{A'}^4}$.
	
	\item \textbf{Dark Matter Velocity Distribution}:
	The dark matter velocity distribution is approximated by a modified Boltzmann distribution. The functions \texttt{dMVelDist} and \texttt{fCross} are the dark matter velocity distribution in the galactic and Earth frames respectively.	The functions \texttt{Normalization} and \texttt{NormalizationChecker} obtain and verify the dark matter velocity distribution normalization $N_0$, \cref{Eqn:f0}.

	\begin{equation}
		\texttt{Normalization} = N_0 = \int_{0}^{v_\text{gal}}
		\left[ \exp \left(\frac{v_\text{gal}^2 - u^2}{k u_0^2}\right) -1 \right]^k \Theta(v_\text{gal} - u)
		\label{eqn:normalization}
	\end{equation}
	with $v_{\text{gal}}=550$ km/s is the escape velocity of the galaxy, $u_0 = 245$ km/s is the average velocity of galactic dark matter, and $k=2.5$ ~\cite{Baratella:2013fya,Choi:2013eda}.
	The dark matter velocity distribution is assumed to be Maxwell--Boltzmann. In the galactic frame, it is $f(u)$ in \cref{Eqn:f0}:
	\begin{equation}
		\texttt{dMVelDist}(u, N_0 = \texttt{Normalization}) = 
		N_0 \left[ \exp \left ( 
		\frac{v^2_{\text{gal}}-u^2}{ku^2_0}\right) -1 \right] ^k \Theta(v_{\text{gal}} - u)
		\label{eqn:DMVelDist}
	\end{equation}
	Upon moving from the galactic frame to the Earth frame, the velocity distribution becomes $f_{\oplus}(u)$ in \cref{Eqn:fcross}:
	\begin{equation}
		\texttt{fCross}(u) = 
		\frac{1}{4} 
\int_{-1}^{1} 
dc_\theta
\int_{-1}^{1}
dc_\phi 
\ f\left[ 
	(u^2 + 
		(V_\odot + V_\oplus c_\gamma c_\phi)^2 + 
		2u(V_\odot + V_\oplus c_\gamma c_\phi)
	c_\theta)^{1/2} 
	\right]\ ,
	\end{equation}
	where $V_{\odot}\approxeq220$ km/s is the velocity of the Sun relative to the galactic center~\cite{Vergados:1998ax}, $V_{\oplus}\approxeq29.8$ km/s is the velocity of the Earth  relative to the Sun~\cite{Gelmini:2000dm}, and $\cos\gamma \approx 0.51$ is the angle of inclination between the orbital planes of the Earth and Sun~\cite{Baratella:2013fya}.
	
	\item \textbf{Kinematics}: These functions define the limits of integration, \cref{Eqn:Eminmax}, in the recoil energy integral of \cref{Eqn:ReducedCapRate}. The lower limit, $E_{\text{min}}$ is the minimum amount of recoil energy that must be transferred from a dark matter particle for it to become gravitationally captured. The upper limit, $E_{\text{max}}$ is the maximum kinematically allowed recoil energy.
	\begin{align}
		\texttt{eMin}(u, m_X) &= E_{\text{min}} = \frac{1}{2} m_X u^2 \\ \texttt{eMax}(N, m_X, i, u) &= E_{\text{max}} = \frac{2\mu^2}{m_N} (u^2 + (v_{\oplus}^2)_i)
		\label{Eqn:eMineMax}
	\end{align}
	The argument \texttt{i} in \texttt{eMax} indexes the radius of Earth at which the escape velocity $v_\oplus$ is calculated. We follow the discretization in the Preliminary Earth Reference Model~\cite{PREM500Paper}.
	We also define the function \texttt{EminEmaxIntersection}, which returns the velocity, $u_{\text{Int}}$, at which $E_{\text{min}}(u_{\text{Int}}) = E_{\text{max}}(u_{\text{Int}})$, \cref{Eq:Eminmax:matching}.
	\begin{equation}
		\texttt{EminEmaxIntersection}(N, m_X, i) = u_{\text{Int}} = \frac{B}{A-B} \left(v_{\oplus}^2\right)_i
		\label{eqn:uInt}
	\end{equation}
	where $A=m_X/2$, $B=2 {\mu^2}/{m_N}$, and $(v_{\oplus})_i=v_{\oplus}(r_i)$ is the escape velocity from Earth at radius $r_i$.

	\item \textbf{Velocity and Energy Integrals}: Here, we define the quantity \texttt{intDuDEr} as the velocity and recoil energy integral inside \cref{Eqn:ReducedCapRate}.
	\begin{equation}
		\texttt{intDuDEr}(N, m_X, m_{A'}, i) = \int_{0}^{u_\text{Int}} du \ u f_\oplus(u) \int_{E_{\text{min}}}^{E_{\text{max}}} dE_R \ \texttt{crossSection}
		\label{eqn:intDuDEr}
	\end{equation}
	\emph{Factors Removed}: $(4\pi)(8\pi \alpha_X \alpha \epsilon^2 m_N Z_N^2)$. The function used for the $\kappa_0$ calculation in \cref{Eqn:Kappa0} is:
	\begin{equation}
		\texttt{intDuDErKappa0}(N, m_X, i) = \int_{0}^{u_\text{Int}} du \ u f_\oplus(u) \int_{E_{\text{min}}}^{E_{\text{max}}} dE_R \ \texttt{crossSectionKappa0}
		\label{eqn:intDuDErKappa0}
	\end{equation}
	\emph{Factors Removed}: $(4\pi)(8\pi \alpha_X \alpha m_N Z_N^2){\varepsilon^2 }/{m_{A'}^4}$.

	\item \textbf{Sum Over Radii}: We define \texttt{sumOverR} which sums \texttt{intDuDEr} over radii of Earth at which the Preliminary Reference Earth Model samples elemental densities~\cite{PREM500Paper}. 
	We calculate the sum in thin shell approximation:
	\begin{equation}
		\frac{4}{3}\pi (r +\Delta r)^3 - \frac{4}{3}\pi r^3 = 4 \pi r^2 \Delta r
		\label{eq:thin:shell}
	\end{equation}
	so that the sum becomes:
	\begin{equation}
		\texttt{sumOverR}(N, m_X, m_{A'}) = \sum_{i} r_i^2 n_N(r_i) \Delta r_i \ \texttt{intDuDEr}
		\label{eqn:sumOverR}
	\end{equation}
	\emph{Factors Removed}: $(4\pi)(4\pi)(8\pi \alpha_X \alpha \epsilon^2 m_N Z_N^2)$.
	$n_N(r)$ is the number density of element $N$ at radius $r$, and the summation index $i$ denotes the $i^\text{th}$ value for radius. The part used to calculate $\kappa_0$, \cref{Eqn:Kappa0}, is:	
	\begin{equation}
		\texttt{sumOverRKappa0}(N, m_X) = \sum_{i} r_i^2 n_N(r_i) \Delta r_i \ \texttt{intDuDErKappa0}
		\label{eqn:sumOverRKappa0}
	\end{equation}
	\emph{Factors Removed}: $(4\pi)(4\pi)(8\pi \alpha_X \alpha m_N Z_N^2){\varepsilon^2}/{m_{A'}^4}$.
	
	
	\item \textbf{Single Element Capture}: We define the capture rate contribution from a single element $N$ to be \texttt{singleElementCap}, \cref{Eqn:ReducedCapRate}. In this function, we restore the previously removed factors $N_\text{SEC} = (4\pi)(4\pi)(8\pi \epsilon^2 \alpha_X \alpha Z_N^2 m_N)$.
	\begin{equation}
		\texttt{singleElementCap}(N, m_X, m_{A'}, \varepsilon, \alpha, \alpha_X) = 
			N_\text{SEC}\,
 			n_X \, \texttt{sumOverR} \ ,
		\label{eqn:singleCap}
	\end{equation}
	where the quantity $n_X = (0.3/m_X) (\text{GeV}/ \text{cm}^3)$ is the local dark matter density \cite{Bovy:2012tw}. Similarly, the part used to calculate $\kappa_0$, \cref{Eqn:Kappa0}, has the previously removed factors restored, 
	$N_{\text{SEC}\kappa_0} = (4\pi)(4\pi)(8\pi \alpha_X \alpha m_N Z_N^2){\varepsilon^2}/{m_{A'}^4}$.:	
	\begin{equation}
		\texttt{singleElementCapKappa0}(N, m_X, \alpha) = 
		N_{\text{SEC}\kappa_0} \,
		n_X \texttt{sumOverRKappa0} \ .
		\label{eqn:singleCapKappa0}
	\end{equation}
	
	
	\item \textbf{Total Capture Rate}: The total capture rate, \cref{Eqn:CapRate}, sums the single-element capture rates over the ten most abundant elements in Earth~\cite{PREM500Paper}.
	\begin{equation}
		\texttt{Ccap}(m_X, m_{A'}, \varepsilon, \alpha, \alpha_X) = C_{\text{cap}} = \sum_N \texttt{singleElementCap}
		\label{eqn:cCap}
	\end{equation}
	We define the functions \texttt{kappa\_0} and \texttt{cCapQuick} to efficiently compute the capture rate in the $(\varepsilon ,m_{A'})$ plane.
	\begin{align}
		\texttt{kappa\_0}(m_X, \alpha) 
			= \kappa_{0}(m_X, \alpha) 
			&= \sum_{N} \texttt{singleElementapKappa0}
		\label{eqn:kappa0}
\\
		\texttt{cCapQuick}(m_X, m_{A'}, \varepsilon, \alpha, \alpha_X) 
			&= \frac{\varepsilon^2 \alpha_X}{m_{A'}^4} \kappa_0
		\label{eqn:cCapQuick} \ .
	\end{align}
\end{itemize}


\subsubsection*{Annihilation Functions}
\begin{itemize}
	\item \textbf{Characteristic Velocity}: This function defines the average velocity of thermalized dark matter particles at the center of the Earth.
	\begin{equation}
		\texttt{v0func}(m_X) = v_0 = \sqrt{2T_{\oplus}/m_X}
	\end{equation}
	where $T_{\oplus} \approx 5700$ K is the temperature at the center of the Earth~\cite{ALFE200291}.

	\item \textbf{Tree-level Annihilation Cross Section}: The annihilation cross section \cref{Eqn:SigmaVTree} without Sommerfeld enhancement is
	\begin{equation}
		\texttt{sigmaVtree}(m_X, m_A, \alpha_X) = (\sigma_{\text{ann}}v)_{\text{tree}} = \frac{\pi \alpha_X^2 [1-m_{A'}^2/m_X^2]^{3/2} }{m_X^2 [1-m_{A'}^2/(2m_X^2)]^2} \ .
	\end{equation}

	\item \textbf{Sommerfeld Enhancement}: The non-perturbative Sommerfeld enhancement \cref{Eqn:Somm} factor that boosts the tree-level annihilation rate in the presence of a long-range force is a function of velocity, dark matter mass, mediator mass, and the dark fine structure constant,
	\begin{equation}
		\texttt{sommerfeld}(v, m_X, m_A, \alpha_X) = S_s = \frac{\pi}{a}\frac{\sinh(2\pi a c)}{\cosh(2\pi a c ) - \cos(2\pi\sqrt{c-a^2c^2)})} \ ,
	\end{equation}
	where $a=v/(2\alpha_X)$ and $c=\alpha_X m_X /(\pi^2 m_{A'})$. This expression is based on the Hulth\'en potential approximation~ \cite{Cassel:2009wt,Feng:2010zp}.
	
	\item \textbf{Thermally Averaged Sommerfeld Enhancement}: The thermally averaged $s$-wave Sommerfeld enhancement \cref{Eqn:SommAvg} is
	\begin{equation}
		\texttt{thermAvgSommerfeld}(m_X, m_A, \alpha_X) = \langle S_S \rangle = \int \frac{d^3v}{(2\pi v_0^2)^{(3/2)}} e^{-\frac{1}{2}v^2/v_0^2 S_s} \ ,
	\end{equation}
	where $v_0$ is the characteristic velocity of dark matter at the center of the Earth.
	
	\item \textbf{Annihilation Rate}: Defines the annihilation rate for thermalized dark matter at the center of the Earth, \cref{Eqn:AnnRate}. 
	\begin{equation}
		\texttt{cAnn}(m_X, (\sigma_{\text{ann}} v)_{\text{tree}}, \langle S_S \rangle = 1) = C_{\text{ann}} = \left[ \frac{G_\text{N}m_X\rho_{\oplus}}{3T_{\oplus}} \right]^{3/2} \langle S_S \rangle (\sigma_{\text{ann}} v)_{\text{tree}}
		\label{eqn:CAnnCalc}
	\end{equation}
	The default thermally-averaged Sommerfeld enhancement is set to unity.
\end{itemize}


\subsubsection*{Signal Functions}

\begin{itemize}
	
	\item \textbf{Equilibrium Time}: The equilibrium time $\tau$, \cref{Eq:equilibrium:time}, is the characteristic time for the dark matter capture and annihilation processes to equilibrate.
	\begin{equation}
		\texttt{tau}(C_{\text{cap}}, C_\text{ann}) = \tau = \sqrt{\frac{1}{C_{\text{cap}} C_\text{ann}}}
	\end{equation}
	
	
	\item \textbf{Contour Function}: Upon specifying a contour level, ${L}$, of $\tau/\tau_{\oplus} = 10^{L}$, this function outputs $\varepsilon$ as a function of $m_{A'}$, \cref{Eqn:EpsilonFunction}.
	\begin{equation}
		\texttt{contourFunction}(m_{A'}, \alpha_X, C_\text{Ann,0}, \langle S_S \rangle, \kappa_0, L) =\log(\varepsilon) = \nonumber
	\end{equation}
	\begin{equation}
		2\log(m_{A'}) -\frac{1}{2}\log(\alpha_X C_\text{ann,0} \langle S_S \rangle) - \frac{1}{2}\log(\kappa_0) - \log(\mathrm{L}  \tau_{\oplus})
	\end{equation}
	where $C_\text{ann,0}$ is the annihilation rate without Sommerfeld effects, given by \cref{eqn:CAnnCalc}, setting $\langle S_S \rangle =1$.
	
	
	\item \textbf{Annihilation Rate}: Defines the rate at which the captured dark matter population decreases due to annihilations, \cref{Eqn:GammaAnnSoln}:
	\begin{equation}
		\texttt{gammaAnn}(C_\text{cap}, C_\text{ann}) = \Gamma_{\text{ann}} = \frac{1}{2} C_{\text{cap}}\tanh^2\left( \frac{\tau_{\oplus}}{\tau}\right)
		\label{gammaAnn}	
	\end{equation}


	\item \textbf{Decay Length}: The decay length, \cref{Eqn:DecayLength}, is the characteristic distance a dark photon produced by dark matter annihilation travels before decaying.
	\begin{equation}
		\texttt{decayLength}(m_X, m_A, \varepsilon, B) = L = R_{\oplus} B \left( \frac{3.6\times 10^{-9}}{\varepsilon} \right)^2 \left( \frac{m_X/m_{A'}}{1000} \right) \left( \frac{\text{GeV}}{m_{A'}} \right)
	\end{equation}
	where $R_{\oplus} = 6370$ km ~\cite{Patrignani:2016xqp} is the radius of earth, $B$ is the branching ratio of $(A' \rightarrow e^+e^-)$, and $\varepsilon$ is the kinetic mixing parameter of the dark and Standard Model photons.
	

	\item \textbf{Decay Parameter}: The decay parameter, \cref{Eqn:EpsilonDecay}, represents the probability that a dark photon produced at the center of Earth propagates to a distance $R_{\oplus}$ within a detector's of effective depth $D$ of the surface. For IceCube, $D\approx 1$ km ~\cite{Ahrens:2002dv}. The quantity $L$ is the characteristic decay length of the dark photon.
	\begin{equation}
		\texttt{epsilonDecay}(L,D=1 \ \text{km}) = \epsilon_{\text{decay}} = e^{-R_{\oplus}/L} - e^{-(R_{\oplus} + D)/L}
		\label{decayParameter}
	\end{equation}


	\item \textbf{Ice Cube Signal Rates}: The function \texttt{iceCubeSignal} gives the number of detectable dark photon decays at IceCube, \cref{Eqn:Nsig}. The parameters $T$ and $A_\text{eff}$ are the observation time in seconds and effective area in \si{cm^2}. For IceCube, $A_{\text{eff}} = 1 \ \text{km}^2 $~\cite{Ahrens:2002dv}.
	\begin{equation}
		\texttt{iceCubSignal}(\Gamma_{\text{ann}},\epsilon_{\text{decay}},T, A_\text{eff}) = N_{\text{sig}} = 2\Gamma_{\text{ann}}\frac{A_{\text{eff}}}{4\pi R_{\oplus}^2}\epsilon_{\text{decay}}T
		\label{nSigEvents}
	\end{equation} 
\end{itemize}


\subsection*{\texttt{brtoe.csv} \label{Sec: brtoe}}
This is csv file contains data about the decay process for $A \rightarrow \ell^+ \ \ell^-$. The data is extracted from~\cite{Buschmann:2015awa}. The first column is the mass of the mediator in GeV and must be labeled \texttt{mA[GeV]}. The second column is the branching ratio of the dark photon $A'$ to a chosen lepton $\ell$ and must be labeled \texttt{BR}. In \cref{tab:brtoe}, we provide the header of \texttt{brtoe.csv}.

\begin{table}[H]
\begin{center}
	\begin{tabular}{l c} 
		\toprule
		\texttt{MA[GeV]} & \texttt{BR} \\ \midrule
		0 & 1 \\
		0.215064 & 1 \\ 
		0.216744 & 0.851 \\ 
		. & . \\
		. & . \\
		. & . \\
		\bottomrule
	\end{tabular}
\end{center}
	\caption{Header of \texttt{brtoe.csv} showing the required column headers.}
	\label{tab:brtoe}
\end{table}


\subsection*{\texttt{AtomicData.py}}
This file defines dictionaries of useful data about elements.
\begin{itemize}
	\item \texttt{elementList}: The 11 primary elements composing Earth in a NumPy array. This list is iterated over when calculating the capture rate, \cref{Eqn:CapRate}.
	
	\item \texttt{atomicNumbers}: The atomic mass in amu of each element in \texttt{elementList} in a Python dictionary. The atomic numbers are accessed as key--value pairs so that \texttt{atomicNumbers[`O16`]} returns $16$.
	
	\item \texttt{nProtons}: The number of protons of each element in \texttt{elementList} in a Python dictionary. 
	
	\item \texttt{coreMassFrac}: The mass fraction of each element in the Earth's core. 
	
	\item \texttt{mantleMassFrac}: The mass fraction of each element in the mantle of Earth.
\end{itemize}


\subsection*{\texttt{Constants.py}}
This file defines all constants that are be used throughout subsequent calculations as global variables~\cite{Patrignani:2016xqp,Baratella:2013fya,Choi:2013eda}. A summary of all constants is provided below in \cref{tab:constants}.

\begin{table}[H]
	\centering
	\begin{tabular}{llll} 
		\toprule
		\textbf{Description} & \textbf{Variable} & \textbf{\texttt{Python}} & \textbf{Value} \\ \midrule
		Speed of light & $c$ & \texttt{c} & $3\times10^{10}$ \ \si{cm.s^{-1}} \\
		Gravitational constant & $G_N$& \texttt{G} & $6.674\times 10^{-8}$ \si{cm^{3}.g^{-1}. s^{-2}}\\
		Sun velocity relative to galactic center & $V_\odot$ & \texttt{V\_dot} & $220.0\times 10^5/c$\\
		Earth velocity relative to Sun & $V_\oplus$ & \texttt{V\_cross} & $29.8\times10^5/c$\\
		Escape velocity of galaxy & $V_\text{gal}$ & \texttt{V\_gal} & $550.0\times 10^5/c$ \\
		Characteristic dark matter velocity & $u_0$ & \texttt{u\_0} & $245.0\times10^5/c $\\
		Velocity distribution power-law index~\cite{Lisanti:2010qx} & $k$ & \texttt{k} & 2.5\\
		Radius of Earth & $R_\oplus$ & \texttt{RCross} & $6.738\times10^8$ \si{cm}\\
		Core-mantle separator& $R_\text{Crit}$ & \texttt{RCrit} & $3.48\times10^8$ \si{cm}\\
		Gravitational constant (natural units) & $G_\text{nat}$ & \texttt{Gnat} & $6.71\times10^{-39}$ \si{GeV^{-2}} \\
		Density at center of Earth & $\rho_\oplus$ & \texttt{rhoCross} & $5.67\times 10^{-17}$ \si{GeV^4} \\
		Temperature at center of Earth & $T_\oplus$ & \texttt{TCross} & $4.9134\times{-10}$ \si{GeV}\\
		Age of Earth & $\tau_{\oplus}$ & \texttt{tauCross} & $4.5\times10^9$ \si{yr}\\
		\bottomrule
	\end{tabular}
	\caption{Table of constants defined in \texttt{Configure/constants.py}.}
	\label{tab:constants}	
\end{table}

\subsection*{\texttt{Conversions.py}}
This file defines useful unit conversions.

\subsection*{\texttt{PlanetData.py}}
This file loads Earth radius and density data from $\texttt{PREM500.csv}$~\cite{PREM500Paper}. Note that we use a modified version of this csv file (included in the \texttt{DarkCapPy} package) to include column headers as needed. 
\begin{itemize}
	\item \texttt{radiusList}: stores 491 values of the radius of Earth. This is used in calculating \texttt{sumOverR}
	
	\item \texttt{densityList}: stores the density of Earth at each radius. This is used in calculating the escape velocity at each radius of Earth.
	
	\item \texttt{deltaRList}: stores 491 values of $\Delta r$. The $i^\text{th}$ element of this list is the difference in radius between the $i^\text{th}$ and $(i+1)^\text{th}$ radii. $\Delta r_i = r_i + r_{i+1}$
	
	\item \texttt{shellMassList}: stores the mass of the $i^\text{th}$ shell according to the thin shell approximation, \cref{eq:thin:shell}. This is used to calculate the enclosed mass.
	
	\item \texttt{enclosedMassList}: stores the mass enclosed up to the $i^\text{th}$ shell. This array is used to calculate the escape velocity.
	
	\item \texttt{escVel2List}: stores the escape velocity, $v_\oplus^2(r_i)$. The $i^\text{th}$ element of this list is the squared escape velocity corresponding to the $i^\text{th}$ radius. This list is used to calculate the lower limit of the recoil energy integration, $E_\text{min}$, \cref{Eqn:Eminmax}.
	
	\item \texttt{numDensityList}: given an input element, this function returns array with the number density of element at each radial distance, $r_i$.
\end{itemize}
From these lists, we generate interpolations of the enclosed mass in grams as a function of radius, the squared escape velocity in natural units as a function of radius $v_\text{esc}^2(r)$, and the density of Earth as a function of radius $n_N(r)$ in cgs units.


\subsection*{\texttt{PREM500.csv}}

Data on the composition of Earth from the Preliminary Reference Earth Model~\cite{PREM500Paper}.  
\texttt{DarkCapPy} only uses the first two columns; these store information about the radius and density of Earth in mks units. The original data file has been modified so the radius and density columns are labeled, \cref{tab:PREM500}.

\begin{table}[H]
	\centering
	\begin{tabular}{l l} 
		\toprule
		\texttt{Radius[m]} & \texttt{Density[kg/m\string^3]} \\ \midrule
		0 & 13088.5 \\ 
		12858 & 13088.46 \\
		25716 & 13088.36 \\ 
		. & . \\
		. & . \\
		. & . \\
		\bottomrule
	\end{tabular}
	\caption{Header of \texttt{PREM500.csv} displaying the columns and required headers}
	\label{tab:PREM500}
\end{table}


\section{Using this Package}

A template Jupyter notebook for performing calculations is available in the \texttt{Template\_Calculation} folder. If \texttt{DarkCapPy} is installed using a package manager, then one may download or view this template notebook directly from GitHub\footnote{\url{https://github.com/agree019/DarkCapPy/blob/master/DarkCapPy/Template_Calculation/DarkCapPy_Template.ipynb}}. This template demonstrates how to run this package for individual parameter points and for a parameter scan. The flow of outputs for a parameter scan is shown in \cref{fig:flowChartAll}. Before starting a calculation, we advise copying the \texttt{Template\_Calculation} folder to a working directory outside of the \texttt{DarkCapPy} package folder within one's Python path.

The Jupyter notebook \texttt{DarkCapPy\_Template.ipynb} is set up to perform  out-of-the-box calculations. Some calculations---in particular those involving a scan over many parameters---may take hours of run time. If the time intensive calculations are interrupted, they must be restarted from the beginning. The template notebook is a guide to using the package to perform time-intensive parameter scans in a way that includes failsafes in the event that the kernel has to be interrupted during a long parameter scan. This package is optimized to perform parameter scans over mediator mass $m_{A'}$ and kinetic mixing $\varepsilon$ for fixed dark matter mass $m_X$.

The template notebook utilizes pandas~\cite{Pandas.2010} to read and write intermediate results to external comma-separated-value (csv) files. These files allow longer calculations to be interrupted and resumed, even if the Jupyter kernel is restarted. This notebook generates 2 csv files, which we refer to as $\texttt{Sommerfeld.csv}$ and $\texttt{Signal.csv}$. It requires input from a third csv file which we refer to as $\texttt{Branch.csv}$.

\tikzstyle{header} = [rectangle, draw]

\tikzstyle{block} = [rectangle, rounded corners, draw, top color = white, bottom color = myblue1!40, text centered, node distance = 5em]

\tikzstyle{line} = [draw, thick, -latex']

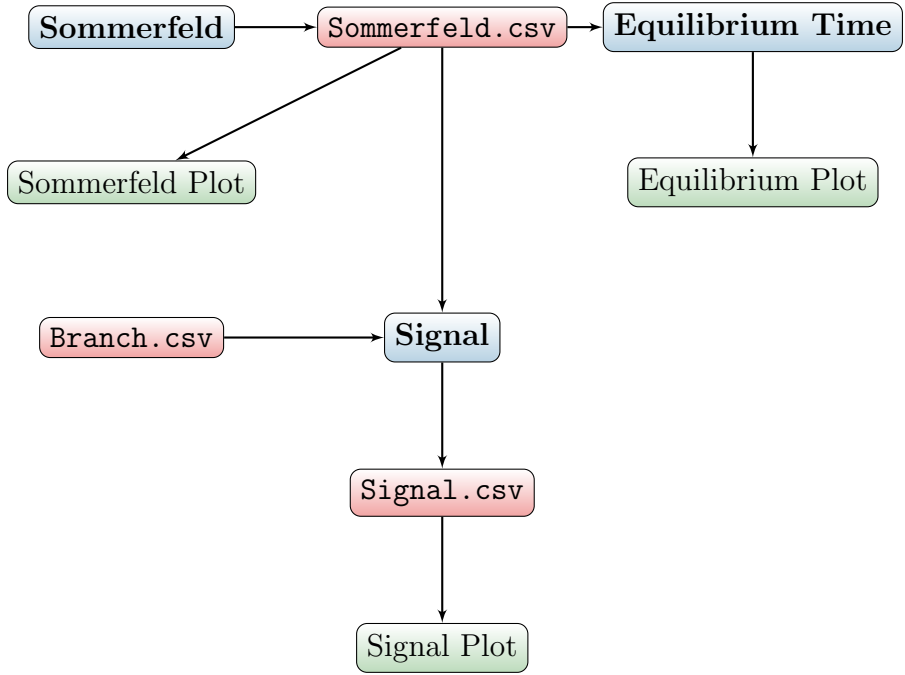
\begin{figure}[H]
	\centering
	\begin{tikzpicture}
	
	\node [block] (Somm) {\bf Sommerfeld};
	
	\node [block, right of=Somm, node distance = 10em ,top color = white, bottom color = myred1!40] (SommFile) {\bf \texttt{Sommerfeld.csv}};
	
	\node [block, below of=Somm, top color = white, bottom color = mygreen1!40] (SommPlot) {Sommerfeld Plot};
	
	\node [block, right of=SommFile, node distance = 10em] (EQ) {\bf Equilibrium Time};
	
	\node [block, below of=EQ, top color = white, bottom color = mygreen1!40] (EQPlot) {Equilibrium Plot};
	
	\node [block, below of=SommFile, node distance = 10em] (Signal) {\bf Signal};
	
	\node [block, below of=Signal, top color = white, bottom color = myred1!40] (SignalFile) {\bf \texttt{Signal.csv}};
	
	\node [block, left of= Signal, node distance = 10em, top color = white, bottom color = myred1!40] (BranchFile) {\bf \texttt{Branch.csv}};
	
	\node [block, below of = SignalFile, top color = white, bottom color = mygreen1!40] (SignalPlot) {Signal Plot};
	
	\path [line] (Somm) -- (SommFile);
	\path [line] (SommFile) -- (EQ);
	\path [line] (EQ) -- (EQPlot);
	\path [line] (SommFile) -- (SommPlot);
	\path [line] (SommFile) -- (Signal);
	\path [line] (BranchFile) -- (Signal);
	\path [line] (Signal) -- (SignalFile);
	\path [line] (SignalFile) -- (SignalPlot);
	\end{tikzpicture}
	\caption{Flowchart outlining the generation of various plots. Blue cells are Sections in \texttt{DarkCapPy\_Template.ipynb}, red cells are input/output csv files, and green cells are visual output of plots.}
	\label{fig:flowChartAll}
\end{figure}

The template notebook is divided into four main sections: Package Test, Sommerfeld, Equilibrium Plots, and IceCube Signal. We provide a brief overview of each section; additional discussion is included in the markdown cells of the Jupyter notebook.

\subsection*{Package Test}

This section tests the package installation by calculating the number of signal events expected at IceCube for a fixed set of parameters---dark matter mass, mediator mass, mediator coupling, and IceCube observation time. It outputs all intermediate results. This section fully demonstrates how to use \texttt{DarkCapPy} for calculations of the `dark Earthshine' scenario of a single set of parameters.

\subsection*{Sommerfeld}

This section creates and populates $\texttt{Sommerfeld.csv}$ with 
\begin{enumerate}
	\item A single value of dark matter mass $m_X$ in GeV
	\item A list mediator masses $m_{A'}$ in GeV
	\item A list of corresponding thermally averaged Sommerfeld enhancements $\langle S_S(m_{A'}) \rangle$
	\item The part of the capture rate that only depends on $m_X$ and $\alpha_X$ in units of GeV$^5$, $\kappa_0$, see \cref{SubSec:Capture} for a discussion of this quantity. 
\end{enumerate}

The user is prompted to input a value of $m_X$ when $\texttt{Sommerfeld.csv}$ is created. A list of mediator masses, $m_{A'}$, are used to generate corresponding values of the thermally averaged Sommerfeld effect, $\langle S_S \rangle$. The quantity $\kappa_0$ depends only on $m_X$ and $\alpha_X$. When $m_X$ is fixed, $\kappa_0$ is uniquely determined by fixing $\alpha_X$ to give the correct dark matter abundance from thermal freeze out. The $\texttt{Sommerfeld.csv}$ file generated from this section is used as input for all later parts of this notebook. The estimated run time for this section is four minutes.

\subsection*{Equilibrium Plots}
This section reads in a completed $\texttt{Sommerfeld.csv}$ and generates plots of the equilibrium time in $(\varepsilon, m_{A'})$ space. The equilibrium time $\tau$ is the time it takes the captured dark matter population to reach a stable value. By default, we plot contour values of $\tau/\tau_\oplus = \{ 10^{-4} ,10^{-2}, 10^0, 10^2, 10^4 \}$ where $\tau_{\oplus} = 10$ Gyr is the age of Earth. 

\subsection*{IceCube Signal}
This section creates and populates $\texttt{Signal.csv}$ file that stores values pertaining to the number of signal events at IceCube. This section is the most time consuming and computation intensive section of this notebook. This is because the typical resolution of these plots requires $\mathcal{O}(10^4)$ calculations.

This section reads in (1) $\texttt{Sommerfeld.csv}$ and (2) $\texttt{Branch.csv}$, and creates $\texttt{Signal.csv}$, which stores the range of mediator masses.

When the calculations are complete, this section also reads in a completed $\texttt{Signal.csv}$ file and plots order-of-magnitude contours of the number of signal events at IceCube $N_\text{sig}$ against mediator mass $m_{A'}$ on the horizontal axis and kinetic mixing parameter $\varepsilon$ on the vertical axis. By default, the contour values are $N_\text{sig} = \{ 1, 10, 100, 1000 \}$. As a benchmark, the calculation rate on a modern laptop is about 140 points / minute.

\section{Hacking this Package}

We intend this package to be modified and changed. Below, we provide a short description of how to modify the parameter ranges, changing the planetary body, changing the mediator and changing the branching ratio.

\subsection{Scaning over Different Parameter Ranges}
W assume the following ranges for $m_{A'}$ and $\varepsilon$:
\begin{equation}
	0.01 \ \text{GeV} \leq m_{A'} \leq 10 \ \text{GeV} \qquad \text{and} \qquad 10^{-11} \leq \varepsilon \leq 10^{-5}
\end{equation}
The lower bound on mediator mass is set by assuming the dark photon decays to electrons. We choose the upper bound so that $(1)$ the dark photons are relativistic when they are created from dark matter annihilations and (2), following the motivation of the study in \cite{Feng:2015hja}, the dark photons are only allowed to decay into electrons. For dark matter mass $\sim 100$ GeV and mediator mass $10 \ \text{MeV} - 10 \ \text{GeV}$, the requirement that the dark photons decay near the surface of Earth, $L \sim R_\oplus$ implies that $\varepsilon \sim 10^{-11} - 10^{-5}$.

Inside \texttt{DarkCapPy\_Template.ipynb} the ``Initialize Parameter Arrays'' of the Sommerfeld section defines the range of mediator mass with the variables \texttt{m\_ALow} and \texttt{m\_AHigh} in GeV. The range defined by these two variables is also used in generating the Signal plots. The range for kinetic mixing is defined in ``Initialize Signal Dataframe'' of the IceCube Signal section with the variables \texttt{epsilon\_Min} and \texttt{epsilon\_Max}.

\subsection{Changing the Planetary Body}
The csv file that is input into \texttt{PlanetData.py} in line 14 defines the planetary body capturing dark matter. For this work, we use \texttt{PREM500.csv}, which stores data about the Earth. Any other data file must have a column storing radius values titled \textbf{Radius[m]} and a column storing the density at each radius titled \textbf{Density[kg/m\string^3]}. To use a different planetary body, change \texttt{PREM500.csv} in line 14 of \texttt{PlanetData.py} to the desired file. Finally, this file must be placed in the same location as \texttt{PlanetData.py}, see \cref{fig:PREMTree} for the file location.
\begin{figure}[H]
	\dirtree{%
	.1 DarkCapPy.
	.2 Configure.
	.3 PlanetData.py.
	.3 PREM500.csv.
}
\caption{Directory tree displaying where \texttt{PREM500.csv} is located and where all planetary csv files must be placed.}
\label{fig:PREMTree}
\end{figure}

\subsection{Changing Branching Ratio}
We provide the branching ratio of dark photons to electrons in the file titled \texttt{brtoe.csv}. The two columns of any other branching ratio csv file must be \textbf{ma[GeV]} and \textbf{BR}. To use a different branching ratio, simply place the csv file in the \texttt{Template\_Calculation/Branching\_Ratio/} folder, see \cref{fig:BranchRatioTree}, and follow the prompts in the ``Initialize Signal.csv'' section to read it in.

\begin{figure}[H]
	\dirtree{%
	.1 DarkCapPy.
	.2 Template\_Calculation.
	.3 Branching\_Ratio.
	.4 brtoe.csv.
}
\caption{Directory tree displaying where branching ratio csv files must be placed.}
\label{fig:BranchRatioTree}
\end{figure}

\subsection{Other modifications}

The package can also be adapted to different mediator spins~\cite{WIP-Chaffey-Green-Tanedo}, capturing bodies~\cite{Feng:2016ijc}, or scattering inelasticity~\cite{Smolinsky:2017fvb}. We leave these modifications for future work.

\section*{Acknowledgements}

We thank Jonathan Feng and Jordan Smolinsky for useful comments and discussions. \textsc{p.t.} and \textsc{a.g.} thank the Keck Institute for Space Sciences for its hospitality during the DaMaSc 2017 symposium where some of this work was developed.
\textsc{p.t.} thanks the Aspen Center for Physics (NSF grant \#1066293) for its hospitality during a period where part of this work was completed and acknowledges support from U.S. Department of Energy under Grant No. DE-SC0008541.


\bibliographystyle{utphys} 	
\bibliography{references}

\end{document}